\documentclass[journal]{IEEEtran}
\ifCLASSINFOpdf
\else
\fi
\usepackage{graphicx}
\usepackage{xcolor}
\usepackage{graphics}
\usepackage{blindtext}
\usepackage{amsmath}
\usepackage{adjustbox}
\usepackage{amsfonts}
\usepackage[]{units}
\usepackage{siunitx}
\usepackage{pdfpages}
\usepackage{amssymb}
\usepackage{fixltx2e}
\usepackage{etoolbox}
\usepackage[noadjust]{cite}
\usepackage{algorithm}
\usepackage{algorithmic}
\usepackage{amsthm,bm}
\usepackage{fancyhdr}
\usepackage{float}

\usepackage{caption} 
\makeatletter
\def\do#1{\patchcmd{#1}{\thepage}{\null}{}{\GenericWarning{}{Could not patch \string#1}}}
\docsvlist{\@oddhead,\@evenhead,\ps@headings,\ps@IEEEtitlepagestyle,\ps@IEEEpeerreviewcoverpagestyle}
\makeatother

\newcommand*\olinea[1]{%
  \vbox{%
    \hrule height 0.5pt
    \kern0.3ex
    \hbox{%
      \kern -0.05em
      \ifmmode#1\else\ensuremath{#1}\fi
      \kern 0.1em
    }
  }
}

\newcommand*\olineb[1]{%
  \vbox{%
    \hrule height 0.5pt
    \kern0.3ex
    \hbox{%
      \kern -0.05em
      \ifmmode#1\else\ensuremath{#1}\fi
      \kern 0em
    }
  }
}

\begin{document}

\bstctlcite{Bibliografia:BSTcontrol}

\title{Constrained Bayesian Active Learning of Interference Channels in Cognitive Radio Networks}

\author{Anestis Tsakmalis, \IEEEmembership{Student Member, IEEE,} Symeon Chatzinotas, \IEEEmembership{Senior Member, IEEE,}\\and Bj\"{o}rn Ottersten, \IEEEmembership{Fellow, IEEE}
\thanks{This work was supported by the European Research Council under the project "AGNOSTIC: Actively Enhanced Cognition based Framework for Design of Complex Systems".}
\thanks{The authors are with the Interdisciplinary Centre for Security, Reliability
and Trust (SnT), University of Luxembourg. Email:\{anestis.tsakmalis, symeon.chatzinotas, bjorn.ottersten\}@uni.lu.}}


\maketitle

\thispagestyle{fancy}


\begin{abstract}

In this paper, a sequential probing method for interference constraint learning is proposed to allow a centralized Cognitive Radio Network (CRN) accessing the frequency band of a Primary User (PU) in an underlay cognitive communication scenario with a designed PU protection specification. The main idea is that the CRN probes the PU and subsequently eavesdrops the reverse PU link to acquire the binary ACK/NACK packet. This feedback indicates whether the probing-induced interference is harmful or not and can be used to learn the PU interference constraint. The cognitive part of this sequential probing process is the selection of the power levels of the Secondary Users (SUs) which aims to learn the PU interference constraint with a minimum number of probing attempts while setting a limit on the number of harmful probing-induced interference events or equivalently of NACK packet observations over a time window. This constrained design problem is studied within the Active Learning (AL) framework and an optimal solution is derived and implemented with a sophisticated, accurate and fast Bayesian Learning method, the Expectation Propagation (EP). The performance of this solution is also demonstrated through numerical simulations and compared with modified versions of AL techniques we developed in earlier work.

\end{abstract}


\begin{IEEEkeywords}

Cognitive Radio Networks, Expectation Propagation, Active Learning, Constrained Dynamic Programming

\end{IEEEkeywords}

\IEEEpeerreviewmaketitle

\section{Introduction}

\IEEEPARstart{U}{}nderlay communication scenarios \cite{biban82} allow the coexistence of a PU and an SU system where SUs may transmit in a PU frequency band as long as the induced interference at the PU is under a certain threshold. This strategy requires some kind of intelligence on the SU side such as cognitive sensing and decision making abilities. These functions basically transform the SU transceivers into powerful and intelligent radio devices which in the telecommunication literature are described as Cognitive Radios (CRs) \cite{biban21, biban97}.

In general, the underlay approach is related to constrained Power Control (PC) or Beamforming (BF) problems where the CR users must intelligently select their transmit power levels or beamforming vectors in order to optimize some operation metric and at the same time satisfy a certain PU interference constraint. Usually, an important piece of these problems, the constraint, is unknown to the CRs, since Channel State Information (CSI) of the interference channels is unavailable at the CRs. Additionally, due to lack of communication between the PU and SU systems, the CRN cannot directly infer the aforementioned parameters, but it must somehow learn these interference channel gains. A common approach for the CRs to overcome this issue is to use the PU reverse link feedback, check how this changes because of the CR operation and thus calculate the SU-to-PU channel gains in a sequential manner. This iterative procedure is clearly a probing scheme which combines carefully selecting the CR transmitting parameters and eavesdropping the PU reverse link feedback. Capturing and exploiting this feedback bridges the gap between the PU and SU systems and enables learning in the CRN.

In the CR literature, the binary ACK/NACK packet of the reverse PU link has been utilized extensively as feedback information. In \cite{biban190}, it is employed to estimate PU receiver maps and in \cite{biban50} to approximate the Lagrange multiplier of the interference constraint in decentralized PC schemes. Moreover, its use has been successful for maximizing or minimizing the power delivered respectively to the SU or PU receiver by adapting the transmit antenna weights in BF scenarios \cite{biban79}. In this paper though, this rudimentary piece of feedback is taken into account only to facilitate learning on the CRN side. Furthermore, a practical and convenient architecture for most CRN scenarios is the CR users to be coordinated by a Cognitive Base Station (CBS) using a dedicated control channel \cite{biban89}. This structure is also chosen here and implicates a centralized network setting which is more applicable than a decentralized CRN where CR users are partially independent and pass messages among each other.

\subsection{Contributions}

Herein, a Constrained Bayesian AL (CBAL) probing method suitable for centrally organized CRNs is demonstrated which rapidly estimates the interference channel gains from multiple SU transmitters to a PU receiver while setting a limit on the number of harmful probing power vectors over a certain time window. This case study assumes that the PU link is operating under a communications protocol where the receiver sends an ACK/NACK packet to the transmitter to acknowledge positively or negatively the receipt of messages. A common practice in the CR regime which is adopted here as well is the CRN to capture this packet from the PU feedback link and exploit it to learn the SU-to-PU channel gains. In this scenario, obtaining this binary feedback takes place in the CBS using a sensing antenna and a PU feedback packet decoder. This piece of information is utilized to implement a sequential probing technique where the SUs constantly adjust their transmit power levels according to CBS directives and monitor whether the ACK/NACK packet changes state.

This intelligent probing design aims to minimize the number of probing attempts which are needed for learning the SU-to-PU channel gains over a time window subject to maintaining the ratio of the harmful probing attempts under a limit. Hence, once the CRN is deployed in the PU system's environment, it may quickly learn the interference channels without severely degrading the PU communication system and then optimize its operation while satisfying the PU interference constraint which depends on the SU-to-PU channel gains. The introduced constraint in this AL process is of practical significance, because it represents the time ratio during which the PU system cannot efficiently operate which is basically an average over time outage constraint, a well defined specification in practical systems. This problem setting is tackled using the Constrained Dynamic Programming (DP) framework. Additionally, exactly because this probing process is sequential, the probing vector design must be implemented fast and accurately at each time step. To achieve this, an advanced Bayesian Learning, the Expectation Propagation (EP) \cite{biban200}, is implemented analytically for the first time to facilitate the AL goal. 

In summary, this paper delivers specifically the following major contributions:
\begin{itemize}
  \item The novel construction of a provenly optimal CBAL method designed for probing the PU and learning fast interference channel gains while maintaining the ratio of harmful probing attempts under a limit.
  \item A computationally cheap, fast and analytical implementation of a sophisticated and accurate Bayesian Learning technique, the EP, suitable for the sequential probing design nature of our problem.
  \item Simulations show fast learning convergence rates for our CBAL method, low required computational burden and most importantly acceptable satisfaction of the harmful interference constraint compared to constrained versions of the Bayesian AL schemes designed in \cite{biban146}.
\end{itemize}
\subsection{Structure} 
The remainder of this paper is structured as follows: Section II reviews in detail prior work related to cognitive learning scenarios using the ACK/NACK feedback of the PU reverse link. Section III provides the system model and the problem formulation. Section IV presents a fast and accurate Bayesian Learning method, the Expectation Propagation, for interference channel gain learning. Section V elaborates on the optimal CBAL probing technique for interference channel gain learning. In Section VI, the simulation results obtained from the application of the proposed technique are shown and compared with the performance of existing methods. Section VII gives the concluding remarks and future work in this topic.


\section{Related work}

In the field of cognitive underlay methods, rudimentary PU feedback has been used for learning purposes in PC and BF scenarios with different assumptions, protocols, system models and constraints. Most commonly, this is acquired by eavesdropping the PU reverse link channel and decoding the PU ACK/NACK packet. The general form of these underlay CR scenarios is the optimization of an SU system metric, such as total CRN throughput, worst SU throughput or SU SINR, subject to QoS constraints for PUs, e.g. SINR, data rate or outage probability \cite{biban82} whose parameters the CRN needs to learn. Hence, these study cases involve learning PU constraints which may be tackled in a centralized manner by a central decision maker or in a decentralized way by each SU individually. Most of the learning techniques are based on a simple iterative scheme of probing the PU system and acquiring the feedback indicating how the PU operation is affected.

In this group of CR learning works, learning the null space of the interference channel matrix in a MIMO underlay cognitive scenario has been tackled by the one-bit null space learning algorithm \cite{biban72}, which essentially is a blind realization of the Cyclic Jacobi Technique. Furthermore, in \cite{biban53}, a binary Spectrum Sensing feedback has been used to enable CRs to apply a Reinforcement Learning procedure, the Q-Learning, to regulate the aggregated interference to the PU. Additionally, in \cite{biban103}, the centralized weighted sum-rate maximization under average SU power and probabilistic PU interference constraints has been considered. In this study, the optimization objective is achieved only after the interference channel gain learning process is terminated, a very common tactic for handling the aforementioned learning and optimization general structure of these problems. In its learning part, the recursive Bayesian estimation is employed by using imperfect CSI feedback which may potentially be as elementary as the binary ACK/NACK packet.

Next, we describe CR learning problems using binary PU feedback which aim at intelligently designing the SU probing attempt in order to learn as fast as possible the unknown constraints of the CR operation. This design rationale is called \textit{Active Learning} (AL) in the ML community and has been approached in many different ways. Initially, the authors of \cite{biban73} proposed a Cutting Plane Method (CPM) based learning algorithm where probing the PU system aims at both learning interference channel matrices and maximizing the SNR at the SU receiver side in an underlay cognitive BF scenario. In \cite{biban146}, we focused only on learning the unknown interference channel gains without optimizing any SU system metric. We proposed an optimal multivariate Bayesian AL method for intelligent probing which incorporates the probability of each feedback being correct and a suboptimal AL method ideal for CRNs with many SUs.

At this point, we need to specify the broader connections of the AL problem setting which led us to the methodology used in this work. AL is tightly connected to a statistical framework called Bayesian Experimental Design \cite{biban199} which in its turn is closely related to the theory of optimal Decision Making. Therefore, researchers from the Decision Making field have exploited a DP approach to sequentially design experiments \cite{biban192}. In \cite{biban195}, the problem of state tracking with active observation control is also tackled in a similar fashion where a Kalman-Like state estimator is developed. Next, AL problems with constraints were developed by the research community which exploited Constrained DP \cite{biban193,biban189} to actively classify human body states with biometric device sensing costs \cite{biban196} and to operate a sensor network with communication costs \cite{biban194}. In this paper, we combine this Constrained DP framework with a sophisticated Bayesian Learning tool, the EP. Moreover, we enhance the accuracy and the speed of the EP by utilizing recent advances in statistics from the econometrics research community \cite{biban191}.


\section{System Model and Problem Formulation}
\begin{figure}[ht!]
\centering
\includegraphics[scale=0.65, trim=10 -15 0 0]{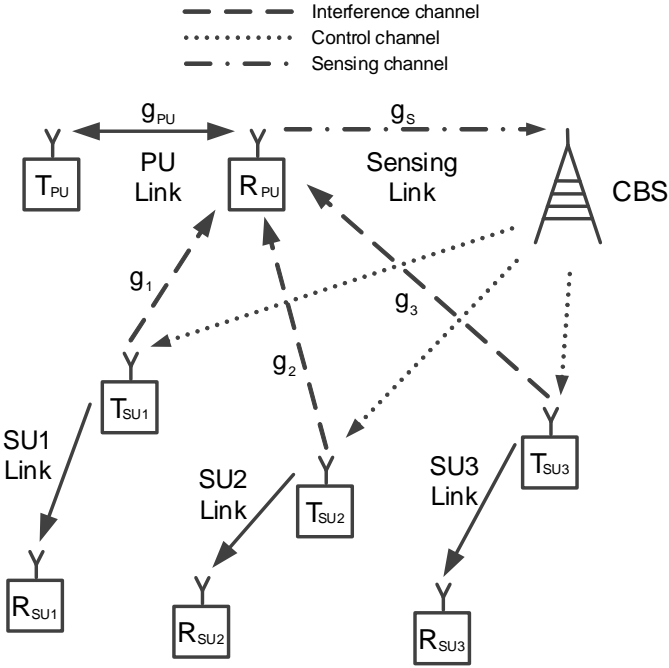}
\caption{The PU system and the CRN}
\label{fig1}
\end{figure}

At first, we describe the system model of our scenario which considers a PU link and $N$ SU links existing in the same frequency band as shown in Fig. \ref{fig1}. A Frequency Division Multiple Access (FDMA) method allows SUs to operate in separate sub-bands of the PU frequency band and without interfering with each other, but still aggregately inducing interference to the PU receiver. The structure of the CRN is a centralized one where the SUs are dictated their power control levels by the CBS using a dedicated control channel. The examined scenarios in this study are considering the PU, the sensing and the unknown interference channels to follow the quasi static block fading model which applies for fixed telecommunication links such as the satellite or the backhauling ones, but not for mobile ones where channels change rapidly. Here, we focus on channel power gains $g$, which are defined as $g=\|c\|^{2}$, where $c$ is the complex channel gain. From this point, we will refer to channel power gains as channel gains.

The interference to the PU link is caused by the transmitter part of each SU link to the receiver of the PU link. Taking into account that the SU links transmit solely in the PU frequency band, the aggregated interference on the PU side is defined as:
\begin{equation}
I_{PU}=\mathbf{g}\;\mathbf{p}^\intercal
\label{eq1}
\end{equation}
where $\mathbf{g}$ is the unknown interference channel gain vector $[g_{1},...,g_{N}]$ with $g_{i}$ being the SU\textsubscript{i}-to-PU interference channel gain and $\mathbf{p}$ is the SU power vector $[p_{1},...,p_{N}]$ with $p_{i}$ being the SU\textsubscript{i} transmit power. These power levels $[p_{1},...,p_{N}]$ are communicated from the CBS to the SUs through the CRN control channel. Additionally, the SINR of the PU is defined as:
\begin{equation}
SINR_{PU}=10\log_{10}\left(\frac{g_{_{PU}}p_{_{PU}}}{I_{PU}+N_{PU}}\right)\mbox{dB}
\label{eq2}
\end{equation}
where $g_{_{PU}}$ is the PU link channel gain, $p_{_{PU}}$ is the PU transmit power and $N_{PU}$ is the PU receiver noise power.

In this study, the CBS is equipped with a secondary omnidirectional antenna only for sensing the signal of the PU reverse link and a module for decoding the binary ACK/NACK feedback. Extracting this binary information $Z$ enables the CRN to detect whether the induced interference to the PU, $I_{PU}$, is harmful or not for the PU data packet reception by the PU receiver. Assuming that $N_{PU}$ and the received power remain the same at the PU receiver side, the minimum required $SINR_{PU}$, $\gamma$, corresponds to a specific unknown maximum allowed $I_{PU}$ value, $I_{th}$, below which an ACK is sent and over which an NACK is transmitted to the PU transmitter. Subsequently, the observed feedback $Z$ is defined as:
\begin{equation}
Z=\left\{
  \begin{array}{cc}
   +1 & \mbox{if $\mathbf{g}\;\mathbf{p}^\intercal\leq I_{th}$}\\
   -1 & \mbox{if $\mathbf{g}\;\mathbf{p}^\intercal> I_{th}$}
  \end{array}
  \right.
\label{eq3}.
\end{equation}
This piece of information will be exploited in the next sections to learn the PU interference constraint determined as:
\begin{equation}
\mathbf{g}\;\mathbf{p}^\intercal \leq I_{th}
\label{eq4}.
\end{equation}

A necessary simplification of the information gained by \eqref{eq3} is that the $g_{i}$ gains normalized to $I_{th}$ are adequate for defining the interference constraint \eqref{eq4}. Therefore, if $\mathbf{h}=\frac{\mathbf{g}}{I_{th}}$, the observed feedback can also be written as:
\begin{equation}
Z=\left\{
  \begin{array}{cc}
   +1 & \mbox{if $\mathbf{h}\;\mathbf{p}^\intercal\leq 1$}\\
   -1 & \mbox{if $\mathbf{h}\;\mathbf{p}^\intercal> 1$}
  \end{array}
  \right.
\label{eq5}
\end{equation}
while the normalized version of \eqref{eq4} is expressed as:
\begin{equation}
\mathbf{h}\;\mathbf{p}^\intercal \leq 1
\label{eq6}.
\end{equation}

In section IV, we elaborate on a sophisticated and computationally fast Bayesian ML method which exploits the observed feedback of \eqref{eq5} to infer \eqref{eq6}. Later, in section V, we propose a CBAL method which achieves learning \eqref{eq6} using the technique described in section IV. The particularity of this CBAL method is that it designs sequentially the SU probing power vectors in order to learn the PU interference constraint with the least probing attempts possible while maintaining a limited number of probing attempts which cause harmful interference.


\section{Bayesian Learning using Expectation Propagation}

In this section, we present a probabilistic way to learn the unknown normalized interference channel gain vector, $\mathbf{h}$, given a set of SU probing power vectors and the corresponding ACK/NACK pieces of feedback. The true value of the unknown normalized interference channel gain vector will be denoted as $\mathbf{h}^{*}$ from here on. These unknown parameters define the constraints \eqref{eq4} and \eqref{eq6} which constitute the PU interference constraint in underlay cognitive scenarios and are also referred to as the interference hyperplane in this work. The data sets of the SU probing power vectors and the ACK/NACK pieces of feedback basically represent the feature vector set and the label set respectively in the ML sense and we demonstrate how to learn the linear classifier, or else the interference hyperplane, denoted by \eqref{eq4} and \eqref{eq6} in the Bayesian way. The reason for following this Bayesian direction will be clearly revealed in the next section, but let us just state here that deriving a pdf for $\mathbf{h}^{*}$ will be proven useful for the AL setting of this paper.

In general, Bayesian ML uses the Bayes rule as the main knowledge extraction tool. To describe in detail the Bayes rule application, first we need to define the feedback, or label, conditional likelihood in this process as the probability of $Z$ conditioned on the unknown parameter $\mathbf{h}^{*}$:
\begin{align}
\Pr [Z|\mathbf{h}=\mathbf{h}^{*},\mathbf{p}^\intercal]=
\left\{
  \begin{array}{cc}
   1 & \mbox{if $Z=+1$ and $\mathbf{h}\;\mathbf{p}^\intercal\leq 1$}\\
   0 & \mbox{if $Z=+1$ and $\mathbf{h}\;\mathbf{p}^\intercal> 1$}\\
   1 & \mbox{if $Z=-1$ and $\mathbf{h}\;\mathbf{p}^\intercal> 1$}\\
   0 & \mbox{if $Z=-1$ and $\mathbf{h}\;\mathbf{p}^\intercal\leq 1$}
  \end{array}
  \right.
\label{eq8}.
\end{align}
This expression is actually a threshold likelihood metric determined by the feedback observation, $Z$, and the power vector $\mathbf{p}$. We may also describe the likelihood function form based on the \textit{version space duality} introduced by Vapnik \cite{biban29}. According to this, when we deal with learning linear classifiers, feature vectors are hyperplanes in the parameter or version space and vice versa. Hence, when a learning procedure tries to estimate the parameters of a hyperplane, the version, it actually tries to localize a point in the parameter or version space. In our problem, the \textit{feature space} corresponds to the power vector space and the \textit{version space} to the $\mathbf{h}$ space. In addition, by combining a power vector, or feature vector, and its respective piece of ACK/NACK feedback, or label, an inequality is obtained which in the $\mathbf{h}$ space, or version space, represents a linear inequality. Therefore, the likelihood function may also be thought of as a halfspace defined by $\mathbf{p}$ and $Z$ or alternatively as a multivariate form of the Heaviside step function in the version space.

Now, let us assume that following $t$ probing attempts, $\mathbf{p}_{0:(t-1)}=\{\mathbf{p}(0),..,\mathbf{p}(t-1)\}$, the CBS has observed $t$ pieces of ACK/NACK feedback, $Z_{0:(t-1)}=\{Z_{0},..,Z_{(t-1)}\}$, which all together constitute the data known until the $(t-1)$ power vector and ACK/NACK feedback pair, $D_{t-1}$. After a new probing power vector $\mathbf{p}(t)$ and a piece of feedback, $Z_{t}$, the $\mathbf{h}$ posterior pdf according to the recursive form of the Bayes rule is expressed as:
\begin{align}
& f_{t+1}(\mathbf{h})=\Pr [\mathbf{h}=\mathbf{h}^{*}|Z_{0:t}, \mathbf{p}_{0:t}]=\Pr [\mathbf{h}=\mathbf{h}^{*}|D_{t}]=\nonumber \\
& \frac{\Pr [Z_{t}|\mathbf{h}=\mathbf{h}^{*},\mathbf{p}(t),D_{t-1}] \; \Pr [\mathbf{h}=\mathbf{h}^{*}|\mathbf{p}(t),D_{t-1}]}{\Pr [Z_{t}|\mathbf{p}(t),D_{t-1}]}
\label{eq9}
\end{align}
which indicates the probability of where $\mathbf{h}^{*}$ lies in the $\mathbf{h}$ space given $D_{t}$. In \eqref{eq9}, we also show the equivalence of the $f_{t+1}(\mathbf{h})$ pdf with the condition $D_{t}$ which represents the knowledge gained until the $t$ step. Here, a necessary remark about the first term of the numerator in \eqref{eq9} must be made which simplifies \eqref{eq9} and which will also help us later. The observation $Z_{t}$ is conditionally independent of the previous observations $Z_{0:(t-1)}$ and probing power vectors $\mathbf{p}_{0:(t-1)}$ given $\mathbf{h}=\mathbf{h}^{*}$ and $\mathbf{p}(t)$ and therefore $\Pr [Z_{t}|\mathbf{h}=\mathbf{h}^{*},\mathbf{p}(t),D_{t-1}]$ can be written as $\Pr [Z_{t}|\mathbf{h}=\mathbf{h}^{*},\mathbf{p}(t)]$ which is basically the likelihood expression in \eqref{eq8}. Moreover, the second term of the numerator, $\Pr [\mathbf{h}=\mathbf{h}^{*}|\mathbf{p}(t),D_{t-1}]$, can be written as $\Pr [\mathbf{h}=\mathbf{h}^{*}|D_{t-1}]$ which is basically the pdf of the previous step, $f_{t}(\mathbf{h})$. This happens because our knowledge about $\mathbf{h}^{*}$ given $Z_{0:(t-1)}$ and $\mathbf{p}_{0:(t-1)}$ does not change by additionally knowing $\mathbf{p}(t)$. After these simplifications the following form of \eqref{eq9} is delivered:
\begin{equation}
f_{t+1}(\mathbf{h})=\frac{\Pr [Z_{t}|\mathbf{h}=\mathbf{h}^{*},\mathbf{p}(t)] \; f_{t}(\mathbf{h})}{\Pr [Z_{t}|\mathbf{p}(t),D_{t-1}]}
\label{eq9a}.
\end{equation}
The denominator term is called the marginal likelihood and even though it is difficult to calculate, it is actually a normalization constant which guarantees that the posterior pdf integrates to 1. Usually, it is computed as the integral of the numerator in \eqref{eq9a} which in our case is an $N$ dimensional integration over the $\mathbf{h}$ region and computationally intractable. A general assumption in Bayesian ML is the prior pdf $f_{0}(\mathbf{h})$ to be a uniform non informative pdf \cite{biban136}, which is the maximum entropy pdf for random variables within a bounded domain and therefore guarantees that no specific value of $\mathbf{h}$ is favored in the beginning of this learning process.

Alternatively, the posterior pdf expressed in \eqref{eq9a} can be written in a non-recursive form as:
\begin{equation}
f_{t+1}(\mathbf{h})=\frac{\prod\limits_{i=0}^{t}\Pr [Z_{i}|\mathbf{h}=\mathbf{h}^{*},\mathbf{p}(i)]}{\prod\limits_{i=0}^{t}\Pr [Z_{i}|\mathbf{p}(i),D_{i-1}]}\;f_{0}(\mathbf{h})
\label{eq9b}
\end{equation}
where again the denominator term is a normalization factor whose computation will be shown unnecessary. The reason we first expressed the posterior pdf in a recursive form is that it will be proven useful in the next section due to the sequential nature of the AL process. Moreover, in Bayesian ML, we should not always take for granted that the posterior pdf is proportional to the \textit{likelihood function product times the prior pdf} which indeed holds for conditionally independent samples. This is the reason why we start from decomposing probabilistically our data set in the Bayes rule expression and first derive its recursive form. More importantly, it is necessary for our AL setting, which relates to Bayesian Experimental Design, to show in detail the conditional independences occurring even when the training samples, here our power probing vectors, are judiciously designed based on previous training samples and their labels.

Now, let us rewrite \eqref{eq9b} in a more compact way in order to focus solely on the likelihood function product and thus approximate it using EP \cite{biban200}. Each likelihood function can be expressed as $l_{i}(\mathbf{h})=\Pr [Z_{i}|\mathbf{h}=\mathbf{h}^{*},\mathbf{p}(i)]$ and hence the likelihood function product of \eqref{eq9b} is now $\prod\limits_{i=0}^{t}l_{i}(\mathbf{h})$. This product is basically a product of halfspace indicator functions and it defines along with $f_{0}(\mathbf{h})$ and the denominator term of \eqref{eq9b}, the marginal likelihood, a uniform pdf with a polyhedral support region. This pdf is not easy to be handled and its statistical properties, like its mean or covariance, are not easily computed. In our previous work \cite{biban146}, this was tackled by using Markov Chain Monte Carlo (MCMC) sampling methods, which are accurate but computationally expensive as the dimensions of the version space increase.

\subsection{The Expectation Propagation algorithm}

In this subsection, we show how to approximate $\prod\limits_{i=0}^{t}l_{i}(\mathbf{h})$ and thus the deriving posterior pdf using EP. The rationale of the EP is to approximate this product by finding an approximation $\tilde{l}_{i}(\mathbf{h})$ for each $l_{i}(\mathbf{h})$. This is done by initializing arbitrarily the likelihood function approximations and iteratively filtering each one of them considering the rest approximations stable. This filtration process is based on minimizing the Kullback–Leibler (KL) divergence of $l_{j}(\mathbf{h})\prod\limits_{i=0,i\neq j}^{t}l_{i}(\mathbf{h})$ and $\tilde{l}_{j}(\mathbf{h})\prod\limits_{i=0,i\neq j}^{t}l_{i}(\mathbf{h})$ and it is performed enough times to ensure that all $\tilde{l}_{i}(\mathbf{h})$ have been corrected sufficiently so that $\prod\limits_{i=0}^{t}\tilde{l}_{i}(\mathbf{h})$ approximates $\prod\limits_{i=0}^{t}l_{i}(\mathbf{h})$ well enough. A detailed algorithmic description of EP is presented in Algo. \ref{alg2}.

\begin{algorithm}
\begin{algorithmic}
\STATE Initialize arbitrarily $\{\tilde{l}_{0}(\mathbf{h}),\tilde{l}_{1}(\mathbf{h}),...,\tilde{l}_{t}(\mathbf{h})\}$
\FOR{$k=1:N_{EP}$}
\FOR{$j=0:t$}
\STATE $\tilde{l}_{j}(\mathbf{h}):=$
\STATE $\arg\min\limits_{\tilde{l}_{j}(\mathbf{h})}KL\left( l_{j}(\mathbf{h})\prod\limits_{i=0,i\neq j}^{t}\tilde{l}_{i}(\mathbf{h}) \parallel \tilde{l}_{j}(\mathbf{h})\prod\limits_{i=0,i\neq j}^{t}\tilde{l}_{i}(\mathbf{h}) \right)$
\ENDFOR
\ENDFOR
\end{algorithmic}
\caption{The Expectation Propagation algorithm}\label{EP}
\label{alg2}
\end{algorithm}

Usually, the outer loop iterations of EP, $N_{EP}$, are chosen to be maximum 5, which is also used in this work. Nevertheless, a more elaborate stopping criterion could be used such as a limit on the KL divergence between the resulting product $\prod\limits_{i=0}^{t}\tilde{l}_{i}(\mathbf{h})$ of one step of the outer loop and the previous one. In Bayesian ML, this sophisticated iterative filtration for likelihood function approximations has proven to be a very accurate method for approximate inference. However, all the existing EP approaches rely on numerical quadratures or independence assumptions between the latent variables to facilitate the computations. Next, we describe in more detail the EP implementation and we show how to tackle \textit{analytically} the KL divergence minimization, the critical step of the EP algorithm, without independence assumptions between the latent variables. This will lead to greater accuracy and faster implementation of this sophisticated tool.

So far, an abstract description of the EP algorithm has been given and its basic principles have been explained. In general, each approximation in the EP algorithm is considered to have the form of a multivariate normal pdf, a strategy which is also followed here. Consequently, the product of multivariate normal pdf's, which appears in the KL divergence minimization step, based on Gaussian identities is also a multivariate normal pdf. More specifically, if $\tilde{l}_{i}(\mathbf{h})=\mathcal{N}(\mathbf{h};\bm{\mu}_{i},\bm{\varSigma}_{i})$ for $i=0,...,t$, where  $\bm{\mu}_{i}$ are the mean row vectors and $\bm{\varSigma}_{i}$ are the covariance matrices, then their product, $\prod\limits_{i=0}^{t}\tilde{l}_{i}(\mathbf{h})$, is an un-normalized multivariate normal pdf proportional to a multivariate normal pdf, $\mathcal{N}(\mathbf{h};\bm{\mu}_{tot},\bm{\varSigma}_{tot})$, where assuming vectors are row vectors:

\begin{equation}
\bm{\varSigma}_{tot}^{-1}=\sum\limits_{i=0}^{t}\bm{\varSigma}_{i}^{-1}
\label{eq10a}
\end{equation}
and 
\begin{equation}
\bm{\mu}_{tot}=\left(\sum\limits_{i=0}^{t}\bm{\mu}_{i}\bm{\varSigma}_{i}^{-1}\right)\bm{\varSigma}_{tot}
\label{eq10b}.
\end{equation}

Hence, the second part of the KL divergence in the core stage of the EP method, $\tilde{l}_{j}(\mathbf{h})\prod\limits_{i=0,i\neq j}^{t}\tilde{l}_{i}(\mathbf{h})$, and the approximation product in the first part, $\prod\limits_{i=0,i\neq j}^{t}\tilde{l}_{i}(\mathbf{h})$, are basically un-normalized multivariate normal pdf's. For notation simplification, $\prod\limits_{i=0,i\neq j}^{t}\tilde{l}_{i}(\mathbf{h})$, which is called the \textit{cavity function}, will be symbolized from now on as $\tilde{l}_{-j}(\mathbf{h})$. Now, as far as the KL divergence minimization is concerned, when Gaussian approximations are used, then this is achieved by \textit{moment matching} \cite{biban136}. A similar theoretical result is also true for all approximations in the exponential family. \textit{Moment matching} means that the two functions whose KL divergence needs to be minimized must have the same moments and since the second function is an un-normalized multivariate normal one, this results to matching the $0_{th}$, $1_{st}$ and $2_{nd}$ moments of the two parts. This basically indicates that the function to be refined in each EP step, $\tilde{l}_{j}(\mathbf{h})$, must be adjusted so that the moments of $\tilde{l}_{j}(\mathbf{h})\;\tilde{l}_{-j}(\mathbf{h})$ are equal to the ones of $l_{j}(\mathbf{h})\;\tilde{l}_{-j}(\mathbf{h})$.

This is the breaking point of the EP algorithm. Calculating the moments of the true likelihood function and the cavity function product could not be implemented so far analytically or in a computationally cheap way. Researchers have tried numerical integration or independence assumptions to simplify the results, but no exact and analytical solution has ever been delivered for basic likelihood function forms. Now, let us examine the function $l_{j}(\mathbf{h})\;\tilde{l}_{-j}(\mathbf{h})$. First, we have already shown that $\tilde{l}_{-j}(\mathbf{h})$ is an un-normalized multivariate normal function and we have described $l_{j}(\mathbf{h})$ as a halfspace indicator function. Thus, $l_{j}(\mathbf{h})\;\tilde{l}_{-j}(\mathbf{h})$ is actually a one-side truncated multivariate Gaussian and what we need is to calculate its $0_{th}$, $1_{st}$ and $2_{nd}$ moments, $q$, $\mathbf{q}$ and $\mathbf{Q}$. To improve the continuity of this manuscript, the analytical moment calculation of a one-side truncated multivariate Gaussian can be found in Appendix \ref{firstAppendix}.

Once these moments are computed, $\tilde{l}_{j}(\mathbf{h})$ is defined using \eqref{eq10a} and \eqref{eq10b} as a multivariate normal pdf with covariance matrix $\bm{\varSigma}_{j}^{-1}=\mathbf{Q}^{-1}-\bm{\varSigma}_{-j}^{-1}$ and mean $\bm{\mu}_{j}=\left( \mathbf{q}\;\mathbf{Q}^{-1}-\bm{\mu}_{-j}\;\bm{\varSigma}_{-j}^{-1} \right)\bm{\varSigma}_{j}$.
We also need to highlight that matching the $0_{th}$ moments does not offer essentially better approximations, because multiplying $\tilde{l}_{j}(\mathbf{h})$ with a constant may lead to unwanted results in this iterative filtration process. Still, we mentioned this earlier as part of the moment matching process for the sake of completeness.

In the end of this subsection, we elaborate on the prior pdf, $f_{0}(\mathbf{h})$. Most commonly, the prior pdf is chosen to represent a prior belief about $\mathbf{h}^{*}$. Moreover, it should also facilitate us computationally in order to have a well defined posterior pdf. If the likelihood functions are approximated with Gaussian ones, then a reasonable choice for $f_{0}(\mathbf{h})$ is also to be Gaussian. Here though, we use another function to show exactly the potential of EP. A closer to reality choice for $f_{0}(\mathbf{h})$ is to define it as a uniform pdf over some bounding box in the $\mathbf{h}$ space. This could represent for example minimum and maximum possible values for $\mathbf{h}^{*}$. Thus, the prior could be described as a hyper-rectangle which can also be written as the product of $2^{N}$ halfspace indicator functions and therefore participate in the EP process.


\section{Constrained Bayesian Active Learning of Interference Channel Gains}

The goal of this paper is to design SU probing power vectors, $\mathbf{p}$, using observations of ACK/NACK feedback, $Z$, in order to learn as fast as possible the unknown normalized interference channel gain vector, $\mathbf{h}^{*}$, while ensuring that the number of probing power vectors causing harmful interference over a time horizon is always below a certain limit. This means that assuming a limited number of $N_{T}$ probing attempts, $\{\mathbf{p}(0),...,\mathbf{p}(N_{T}-1)\}$ and their corresponding pieces of feedback, $\{Z_{0},...,Z_{N_{T}-1}\}$, we wish to minimize the uncertainty of our knowledge about $\mathbf{h}^{*}$, formally represented by the entropy of $f_{N_{T}}(\mathbf{h})$, subject to maintaining the sum of $Z_{t}=-1$, where $t=0,...,N_{T}-1$, below a threshold and which is equivalent to controlling the sum of $Z_{t}=+1$, where $t=0,...,N_{T}-1$, above a corresponding limit. This practical constraint is essential for the PU system operation, since the actual deterioration of its link does not depend on the total or average amount of interference over time caused by the CRN, but on the time ratio during which harmful interference occurs because of SU probing attempts.

\begin{figure}[ht!]
\centering
\includegraphics[scale=0.484, trim=0 -20 0 0]{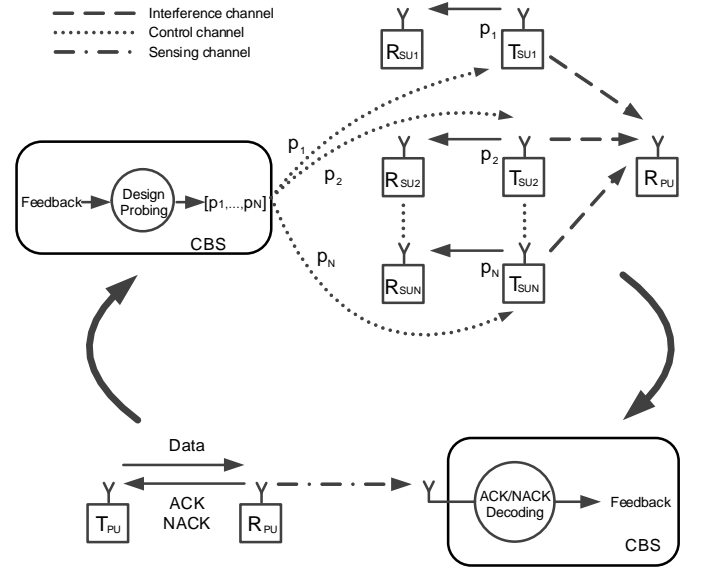}
\caption{The Active Learning probing scheme}
\label{fig12}
\end{figure}

In the previous section, we showed the recursive Bayesian update \eqref{eq9a} which modifies our knowledge about $\mathbf{h}^{*}$ step by step. This will be our main tool for handling the iterative nature of this proactive probing strategy.  In Fig. \ref{fig12}, we may also see how this repeated probe designing and probing scheme is carried out in our cognitive scenario where the CRN designs its probing power vector and probes the PU and subsequently monitors the ACK/NACK feedback sent by the PU receiver in order to infer the interference hyperplane and then repeats the same process.

\subsection{The DP formulation of the Constrained Bayesian AL problem}

Next, we investigate the optimal design policy of a SU probing power vector, which represents a hyperplane in the $\mathbf{h}$ space, that should be chosen in each step of this recursive Bayesian estimation process in order to optimally reduce the posterior pdf entropy after $N_{T}$ probing power vectors, $\{\mathbf{p}(0),...,\mathbf{p}(N_{T}-1)\}$, with their corresponding pieces of feedback, $Z_{0:(N_{T}-1)}$, subject to $\frac{1}{N_{T}}\sum\limits_{t=0}^{N_{T}-1} \mathbf{1}_{\{Z_{t}=-1\}} \leq \alpha'$ or $\frac{1}{N_{T}}\sum\limits_{t=0}^{N_{T}-1} \mathbf{1}_{\{Z_{t}=+1\}} \geq \alpha$, where $\mathbf{1}_{\{..\}}$ is the indicator function, $\alpha'$ is the harmful interference time ratio, $\alpha$ is the harmless interference or protection time ratio during which the PU link operation must remain undisrupted and $\alpha=1-\alpha'$. From here on, we employ the protection time ratio $\alpha$ for the formulation of our problem. The constraint can also be written as $\sum\limits_{t=0}^{N_{T}-1} Z_{t} \geq (2\alpha-1)N_{T}$. This multistage constrained optimization problem can be expressed in the spirit of DP \cite{biban106} as finding the optimal probing rule that maps $\{f_{0},..,f_{N_{T}-1}\}$ to $\{\mathbf{p}(0),..,\mathbf{p}(N_{T}-1)\}$ in order to achieve the maximum average entropy reduction from the $f_{0}(\mathbf{h})$ to the $f_{N_{T}}(\mathbf{h})$ pdf subject to the aforementioned constraint. In a formal manner, we seek the optimal probing design policy $\pi^{*}_{0:(N_{T}-1)}=\{\mathbf{p}(0)=\mu^{*}(f_{0}),..,\mathbf{p}(N_{T}-1)=\mu^{*}(f_{N_{T}-1})\}$ which solves the following constrained optimization problem over all possible feedback sequences derived by this policy:

\begin{subequations}
 \label{eq12:optim}
 \begin{align}
 & \underset{\pi}{\text{max}}
 & & E^{\pi}[\mathcal{H}(f_{0})-\mathcal{H}(f_{N_{T}})|\mathbf{p}(N_{T}-1),D_{N_{T}-2}] \label{eq12:a} \\
 & \text{s.t.}
 & & E^{\pi}\left[\sum\limits_{t=0}^{N_{T}-1} Z_{t}|\mathbf{p}(N_{T}-1),D_{N_{T}-2} \right] \geq (2\alpha-1)N_{T} \label{eq12:b}
 \end{align}
\end{subequations}
where $\mathcal{H}$ is the entropy operator of a pdf. The objective function of \eqref{eq12:optim} which is the conditional expectation of the information gain of an arbitrary policy $\pi$ can also be expressed in an additive form:
\begin{align}
& E^{\pi}[\mathcal{H}(f_{0})-\mathcal{H}(f_{N_{T}})|\mathbf{p}(N_{T}-1),D_{N_{T}-2}]=\nonumber \\
& E^{\pi}[\mathcal{H}(f_{0})-\mathcal{H}(f_{1})|\mathbf{p}(0)+...\nonumber \\
&+E^{\pi}[\mathcal{H}(f_{k-1})-\mathcal{H}(f_{k})|\mathbf{p}(k-1),D_{k-2}+...\nonumber \\
&+E^{\pi}[\mathcal{H}(f_{N_{T}-1})-\mathcal{H}(f_{N_{T}})|\mathbf{p}(N_{T}-1),D_{N_{T}-2}]...]
\label{eq13}
\end{align}
where we added and subtracted all the entropy terms of the intermediate pdf's to form an additive gain over time and similarly the left part of the constraint of \eqref{eq12:optim} can be written as:
\begin{align}
& E^{\pi}\left[\sum\limits_{t=0}^{N_{T}-1} Z_{t}|\mathbf{p}(N_{T}-1),D_{N_{T}-2} \right]=\nonumber \\
& E^{\pi}[Z_{0}|\mathbf{p}(0)+...+E^{\pi}[Z_{k-1}|\mathbf{p}(k-1),D_{k-2}+...\nonumber \\
&+E^{\pi}[Z_{N_{T}-1}|\mathbf{p}(N_{T}-1),D_{N_{T}-2}]...]
\label{eq14}.
\end{align}

After we invert the entropy subtractions, in order to reform the optimization problem into a minimization one, and move the left part of \eqref{eq12:b} on the right side, we create the Lagrangian of this multistage problem as:
\begin{align}
& J^{\lambda}_{0:(N_{T}-1)}=E^{\pi}[\mathcal{H}(f_{1})-\mathcal{H}(f_{0})-\lambda Z_{0}|\mathbf{p}(0)+...\nonumber \\
&+E^{\pi}[\mathcal{H}(f_{k})-\mathcal{H}(f_{k-1})-\lambda Z_{k-1}|\mathbf{p}(k-1),D_{k-2}+...\nonumber \\
&+E^{\pi}[\mathcal{H}(f_{N_{T}})-\mathcal{H}(f_{N_{T}-1})-\lambda Z_{N_{T}-1}|\mathbf{p}(N_{T}-1),D_{N_{T}-2}]...]+\nonumber \\
&+\lambda (2\alpha -1)N_{T}
\label{eq15}
\end{align}
where $\lambda$ is the KKT multiplier related to \eqref{eq12:b} and which has to be non-negative. Now, we need to minimize $J^{\lambda}_{0:(N_{T}-1)}$ for an abstract $\lambda$ and we can do so without including the last term $\lambda (2\alpha -1)N_{T}$, since it is independent of the policy $\pi$. The new form of the Lagrangian will thus be $\Lambda^{\lambda}_{0:(N_{T}-1)}=J^{\lambda}_{0:(N_{T}-1)}-\lambda (2\alpha -1)N_{T}$. Additionally, to bring our problem closer to the DP formulation, we define the subtail problem Lagrangian or Lagrangian-to-go, $\Lambda^{\lambda}_{k:(N_{T}-1)}$, as:
\begin{align}
& \Lambda^{\lambda}_{k:(N_{T}-1)}=\nonumber \\
& E^{\pi}[\mathcal{H}(f_{k+1})-\mathcal{H}(f_{k})-\lambda Z_{k}|\mathbf{p}(k),D_{k-1}+...\nonumber \\
& +E^{\pi}[\mathcal{H}(f_{N_{T}})-\mathcal{H}(f_{N_{T}-1})-\lambda Z_{N_{T}-1}|\mathbf{p}(N_{T}-1),D_{N_{T}-2}]...]
\label{eq16}.
\end{align}
and we denote its minimum value as $\Lambda^{*\lambda}_{k:(N_{T}-1)}$. By employing the principle of optimality, we have:
\begin{align}
& \Lambda^{*\lambda}_{k:(N_{T}-1)}=\nonumber \\
& \underset{\pi}{\text{min}}\;E^{\pi}\left[ \mathcal{H}(f_{k+1})-\mathcal{H}(f_{k})-\lambda Z_{k}|\mathbf{p}(k),D_{k-1}+\Lambda^{*\lambda}_{(k+1):(N_{T}-1)} \right]
\label{eq17}
\end{align}
and based on this, we may proceed with the backward induction logic of DP.

Before we continue though with the DP solution of our constrained multistage problem, let us first redefine the multivariate cumulative distribution function (cdf) in a more "natural" than the usual way. Assuming a multivariate pdf $f$ in $S \subseteq \mathbb{R}^N$ and a vector $\mathbf{x}=[x_{1},...,x_{N}]$, usually its cdf $F$ is defined as $F(\mathbf{x})=\Pr [X_{1}\leq x_{1},...,X_{N}\leq x_{N}]$ which is the joint probability of its components $X_{1},...,X_{N}$, that are scalar valued random variables, being less or equal than the values $x_{1},...,x_{N}$ respectively. Nevertheless, this definition is not geometrically smooth and commonly used just because it is easy to be computed in case of independent $\mathbf{x}$ components. Here, we describe it more strictly and not just by using a "box limit"-like definition. Assuming a hyperplane in $\mathbb{R}^n$, $\mathbf{x}\;\mathbf{w}^\intercal=1$, we alternatively determine the cdf $C$ of a multivariate pdf $f$ as:
\begin{equation}
C(\mathbf{w})=\Pr [\mathbf{x}\;\mathbf{w}^\intercal \leq 1]=\int\limits_{\mathbf{x}\;\mathbf{w}^\intercal \leq 1} f(\mathbf{x})\;dV_{\mathbf{x}}
\label{eq19}.
\end{equation}

For our case study, this means that the posterior cdf after the $(t-1)$ step, $C_{t}(\mathbf{p})$, is expressed as:
\begin{equation}
C_{t}(\mathbf{p})=\Pr [\mathbf{h}\;\mathbf{p}^\intercal \leq 1|\mathbf{h}=\mathbf{h}^{*},D_{t-1}]=\int\limits_{\;\mathbf{h}\;\mathbf{p}^\intercal \leq 1} f_{t}(\mathbf{h})\;dV_{\mathbf{h}}
\label{eq20}
\end{equation}
and the support region of $f_{t}(\mathbf{h})$ is limited to the positive orthant of the $\mathbf{h}$ space, $\mathbb{R}^{N}_{+}$, because the interference channel gains can only have non negative values.

Further on, we elaborate on the marginal likelihood of \eqref{eq9a}. In the event of $Z_{t}=+1$, the conditional probability $\Pr [Z_{t}|\mathbf{p}(t),D_{t-1}]$ can also be written according to the Bayes sum rule, the product rule and the conditional independences from Section IV as in \eqref{eq21}.
\begin{figure*}
\begin{gather}
\Pr [Z_{t}=+1|\mathbf{p}(t),D_{t-1}]=\int\limits_{\mathbb{R}^{N}_{+}} \Pr [Z_{t}=+1,\mathbf{h}=\mathbf{h}^{*}|\mathbf{p}(t),D_{t-1}]\;dV_{\mathbf{h}}= \nonumber \\
 \int\limits_{\mathbb{R}^{N}_{+}} \frac{\Pr [Z_{t}=+1|\mathbf{h}=\mathbf{h}^{*},\mathbf{p}(t),D_{t-1}] \; \Pr [\mathbf{h}=\mathbf{h}^{*},\mathbf{p}(t),D_{t-1}]}{\Pr [\mathbf{p}(t),D_{t-1}]} dV_{\mathbf{h}}=\nonumber \\
 \int\limits_{\mathbb{R}^{N}_{+}} \Pr [Z_{t}=+1|\mathbf{h}=\mathbf{h}^{*},\mathbf{p}(t),D_{t-1}] \Pr [\mathbf{h}=\mathbf{h}^{*}|\mathbf{p}(t),D_{t-1}] dV_{\mathbf{h}}=\nonumber \\
 \int\limits_{\mathbf{h}\;\mathbf{p}^\intercal \leq 1} \Pr [Z_{t}=+1|\mathbf{h}=\mathbf{h}^{*},\mathbf{p}(t),D_{t-1}] \Pr [\mathbf{h}=\mathbf{h}^{*}|\mathbf{p}(t),D_{t-1}] dV_{\mathbf{h}}+\nonumber \\
 \int\limits_{\mathbf{h}\;\mathbf{p}^\intercal > 1} \Pr [Z_{t}=+1|\mathbf{h}=\mathbf{h}^{*},\mathbf{p}(t),D_{t-1}] \Pr [\mathbf{h}=\mathbf{h}^{*}|\mathbf{p}(t),D_{t-1}] dV_{\mathbf{h}}=\nonumber \\
 \int\limits_{\mathbf{h}\;\mathbf{p}^\intercal \leq 1} \Pr [Z_{t}=+1|\mathbf{h}=\mathbf{h}^{*},\mathbf{p}(t),D_{t-1}] \Pr [\mathbf{h}=\mathbf{h}^{*}|D_{t-1}] dV_{\mathbf{h}}+\nonumber \\
 \int\limits_{\mathbf{h}\;\mathbf{p}^\intercal > 1} \Pr [Z_{t}=+1|\mathbf{h}=\mathbf{h}^{*},\mathbf{p}(t),D_{t-1}] \Pr [\mathbf{h}=\mathbf{h}^{*}|D_{t-1}] dV_{\mathbf{h}}=\nonumber \\
 \int\limits_{\mathbf{h}\;\mathbf{p}^\intercal \leq 1} \Pr [Z_{t}=+1|\mathbf{h}=\mathbf{h}^{*},\mathbf{p}(t),D_{t-1}] \; f_{t}(\mathbf{h}) dV_{\mathbf{h}}+\int\limits_{\mathbf{h}\;\mathbf{p}^\intercal > 1} \Pr [Z_{t}=+1|\mathbf{h}=\mathbf{h}^{*},\mathbf{p}(t),D_{t-1}] \; f_{t}(\mathbf{h}) dV_{\mathbf{h}}=\nonumber \\
 \int\limits_{\mathbf{h}\;\mathbf{p}^\intercal \leq 1} \Pr [Z_{t}=+1|\mathbf{h}=\mathbf{h}^{*},\mathbf{p}(t)] \; f_{t}(\mathbf{h}) dV_{\mathbf{h}}+\int\limits_{\mathbf{h}\;\mathbf{p}^\intercal > 1} \Pr [Z_{t}=+1|\mathbf{h}=\mathbf{h}^{*},\mathbf{p}(t)] \; f_{t}(\mathbf{h}) dV_{\mathbf{h}}=\nonumber \\
 \int\limits_{\mathbf{h}\;\mathbf{p}^\intercal \leq 1} f_{t}(\mathbf{h}) dV_{\mathbf{h}}= C_{t}(\mathbf{p}(t))
\label{eq21}
\end{gather}
\hrule
\end{figure*}

A similar expression can also be derived for the $Z_{t}=-1$ event:
\begin{equation}
\Pr [Z_{t}=-1|\mathbf{p}(t),D_{t-1}]=1-C_{t}(\mathbf{p}(t))
\label{eq23}.
\end{equation}

Moving on with our DP solution, we apply the backward induction logic of DP and first solve $\underset{\pi}{\text{min}}\;E^{\pi}\left[ \Lambda^{\lambda}_{(N_{T}-1):(N_{T}-1)} \right]$ which is equivalent to:
\begin{align}
\underset{\mathbf{p}(N_{T}-1)}{\text{min}}\; & E^{\pi}[ \mathcal{H}(f_{N_{T}})-\mathcal{H}(f_{N_{T}-1})- \nonumber \\ 
& -\lambda Z_{N_{T}-1}|\mathbf{p}(N_{T}-1),D_{N_{T}-2} ]
\label{eq24}.
\end{align}
Now, let us first evaluate the term $E^{\pi}[ \mathcal{H}(f_{N_{T}})-\mathcal{H}(f_{N_{T}-1})-\lambda Z_{N_{T}-1}|\mathbf{p}(N_{T}-1),D_{N_{T}-2} ]$, where $E^{\pi}[.]$ is basically the expectation over the two possible observations $Z_{N_{T}-1}=+1$ and $Z_{N_{T}-1}=-1$, by using \eqref{eq9a} and the equivalence of the conditions $D_{N_{T}-2}$ and $f_{N_{T}-1}$:
\begin{align}
& E^{\pi}[ \mathcal{H}(f_{N_{T}})-\mathcal{H}(f_{N_{T}-1})-\lambda Z_{N_{T}-1}|\mathbf{p}(N_{T}-1),f_{N_{T}-1} ]=\nonumber \\
& E^{\pi} \left[ E_{\mathbf{h}}\left[-\log(f_{N_{T}-1})\right]\right] -E^{\pi} \left[ E_{\mathbf{h}}\left[-\log(f_{N_{T}-1})\right]\right]+\nonumber \\
& +E^{\pi} [E_{\mathbf{h}}\left[ -\log(\Pr [Z_{N_{T}-1}|\mathbf{h}=\mathbf{h}^{*},\mathbf{p}(N_{T}-1)])\right]|\nonumber \\
& |\mathbf{p}(N_{T}-1),f_{N_{T}-1}]-\nonumber \\
& -E^{\pi} [E_{\mathbf{h}}\left[-\log(\Pr [Z_{N_{T}-1}|\mathbf{p}(N_{T}-1),f_{N_{T}-1}])\right]|\nonumber \\
& |\mathbf{p}(N_{T}-1),f_{N_{T}-1}]-\nonumber \\
& -\lambda E^{\pi}[ Z_{N_{T}-1}|\mathbf{p}(N_{T}-1),f_{N_{T}-1} ]
\label{eq27}.
\end{align}
The last three remaining terms can be further processed. With the help of \eqref{eq8} for $\Pr [Z_{N_{T}-1}|\mathbf{h}=\mathbf{h}^{*},\mathbf{p}(N_{T}-1)]$, the third term can be analyzed as:
\begin{align}
& E^{\pi} [E_{\mathbf{h}}\left[ -\log(\Pr [Z_{N_{T}-1}|\mathbf{h}=\mathbf{h}^{*},\mathbf{p}(N_{T}-1)])\right]|\nonumber \\
& |\mathbf{p}(N_{T}-1),f_{N_{T}-1}]=\nonumber \\
& E^{\pi} \left[E_{\mathbf{h}}\left[ -\log(\Pr [Z_{N_{T}-1}|\mathbf{h}=\mathbf{h}^{*},\mathbf{p}(N_{T}-1)])\right]|f_{N_{T}-1} \right]=\nonumber \\
& E^{\pi} \left[ -\log(\Pr [Z_{N_{T}-1}|\mathbf{h}=\mathbf{h}^{*},\mathbf{p}(N_{T}-1)])\right]=0
\label{eq28}
\end{align}
where we exploited the fact that $Z_{N_{T}-1}$ does not depend on $f_{N_{T}-1}$ given $\mathbf{h}=\mathbf{h}^{*}$ and $\mathbf{p}(N_{T}-1)$. Additionally, by using \eqref{eq21} and \eqref{eq23} which again lead us to omit $E_{\mathbf{h}}$, since $\Pr [Z_{N_{T}-1}|\mathbf{p}(N_{T}-1),f_{N_{T}-1}]$ is stable over the $\mathbf{h}$ domain, the fourth term becomes:
\begin{align}
& E^{\pi} [E_{\mathbf{h}}\left[-\log(\Pr [Z_{N_{T}-1}|\mathbf{p}(N_{T}-1),f_{N_{T}-1}])\right]|\nonumber \\
& |\mathbf{p}(N_{T}-1),f_{N_{T}-1}]=\nonumber \\
& E^{\pi} \left[-\log(\Pr [Z_{N_{T}-1}|\mathbf{p}(N_{T}-1),f_{N_{T}-1}]) \right]=\nonumber \\
& -C_{N_{T}-1}(\mathbf{p}(N_{T}-1))\log(C_{N_{T}-1}(\mathbf{p}(N_{T}-1)))-\nonumber \\
& -(1-C_{N_{T}-1}(\mathbf{p}(N_{T}-1)))\log((1-C_{N_{T}-1}(\mathbf{p}(N_{T}-1))))
\label{eq29}.
\end{align}
Finally, we elaborate on the fifth term:
\begin{align}
& \lambda E^{\pi}[ Z_{N_{T}-1}|\mathbf{p}(N_{T}-1),f_{N_{T}-1} ]=\nonumber \\
& \lambda [(+1) \Pr[Z_{N_{T}-1}=+1|\mathbf{p}(N_{T}-1),f_{N_{T}-1}] +\nonumber \\
& + (-1) \Pr[Z_{N_{T}-1}=-1|\mathbf{p}(N_{T}-1),f_{N_{T}-1}]]=\nonumber \\
& \lambda[C_{N_{T}-1}(\mathbf{p}(N_{T}-1))-(1-C_{N_{T}-1}(\mathbf{p}(N_{T}-1)))]
\label{eq30}.
\end{align}

We observe that minimizing \eqref{eq27} using \eqref{eq28}, \eqref{eq29} and \eqref{eq30} over $\mathbf{p}(N_{T}-1)$ is equivalent to minimizing \eqref{eq27} over $C_{N_{T}-1}$, since the term $\mathbf{p}(N_{T}-1)$ appears only inside $C_{N_{T}-1}(.)$. Consequently, this results to the following problem where we include \eqref{eq28}, \eqref{eq29} and \eqref{eq30} in \eqref{eq27} and simplify the notation for the sake of space with the help of $C=C_{N_{T}-1}(\mathbf{p}(N_{T}-1))$:
\begin{align}
& \Lambda^{\lambda}_{(N_{T}-1):(N_{T}-1)}=\nonumber \\
& C\log(C)+(1-C)\log(1-C)-\lambda (2C-1)
\label{eq31}
\end{align}
and thus \eqref{eq24} becomes:
\begin{equation}
\underset{C}{\text{min}}\; [C\log(C)+(1-C)\log(1-C)-\lambda (2C-1)]
\label{eq32}.
\end{equation}
Solving \eqref{eq32} by imposing $\frac{\partial \Lambda^{\lambda}_{(N_{T}-1):(N_{T}-1)}}{\partial C}=0$ results to the value of $C=\frac{e^{2\lambda}}{1+e^{2\lambda}}$ which delivers $ \Lambda^{*\lambda}_{(N_{T}-1):(N_{T}-1)}=\lambda-\log(1+e^{2\lambda})$. We notice that this minimum value of the Lagrangian-to-go $\Lambda^{\lambda}_{(N_{T}-1):(N_{T}-1)}$ is a constant value and independent of the time step. This allows us to state that by moving backwards in time at the $(k+1)$ time step, the accumulated constant values of the of the Lagrangian's-to-go yield the following:
\begin{equation}
\Lambda^{*\lambda}_{(k+1):(N_{T}-1)}=\left( (N_{T}-1)-(k+1)+1 \right)\left(\lambda-\log(1+e^{2\lambda})\right)
\label{eq33}.
\end{equation}
Proceeding with our DP solution, we now solve \eqref{eq17} using the same procedure as before and we obtain that:
\begin{equation}
\Lambda^{*\lambda}_{k:(N_{T}-1)}=\left( N_{T}-k \right)\left(\lambda-\log(1+e^{2\lambda})\right)
\label{eq34}
\end{equation}
which for $k=0$ gives $\Lambda^{*\lambda}_{0:(N_{T}-1)}=N_{T}\left(\lambda-\log(1+e^{2\lambda})\right)$. Consequently, the dual function $q(\lambda)$ of \eqref{eq12:optim}, which is always concave, is defined as:
\begin{align}
& q(\lambda)=J^{*\lambda}_{0:(N_{T}-1)}=\Lambda^{*\lambda}_{0:(N_{T}-1)}+\lambda (2\alpha -1)N_{T}= \nonumber \\
& =N_{T}\left(\lambda-\log(1+e^{2\lambda})\right)+\lambda (2\alpha -1)N_{T}
\label{eq35}
\end{align}
which enables us to rewrite \eqref{eq12:optim} as:
\begin{subequations}
 \label{eq36:optim}
 \begin{alignat}{2}
 & \underset{\lambda}{\text{max}} & q(\lambda) \label{eq136:a} \\
 & \text{s.t.} & \lambda \geq 0 \label{eq36:b}
 \end{alignat}
\end{subequations}
and solve this by imposing $\frac{\partial q(\lambda)}{\partial \lambda}=0$ which delivers $\lambda^{*}=0.5\log (\frac{\alpha}{1-\alpha})$. For $\alpha \geq 0.5$, which is the lower reasonable limit of the time ratio during which the CRN probes protectively to the PU system, we always have $\lambda^{*}>0$ and therefore the constraint \eqref{eq12:b} is active because of the complementary slackness condition. Finally, we conclude by using $\lambda^{*}$ that the optimal probing design policy must satisfy $C_{t}(\mathbf{p}(t))=\alpha$ or equivalently $\mathbf{p}(t)=\mu^{*}(f_{t}(\mathbf{h}))=C_{t}^{-1}(\alpha)$ for every time step and for this reason $\mathbf{p}(0)=C_{0}^{-1}(\alpha)$.

At this point, we must emphasize some aspects of the optimal policy. This constrained DP problem must somehow take into account the actual obtained pieces of feedback and not just the expected ones derived from the probabilistic formulation of our problem. This is similar to inventory control problems with stock constraints where we may know probabilistically the product demands over time, but we also need to include into the inventory control the actual demands already arrived before each time step. This means that the protection time ratio $\alpha$ should be adapted to the past feedback observations.

Now, let us take a closer look to the optimal policy at some arbitrary time step $k$. The multistage optimization problem in time step $k$ has a form similar to the one of \eqref{eq12:optim}, only that this time we are interested in maximizing the information gain in the remaining steps and still maintaining the overall violation constraint: 

\begin{subequations}
 \label{eq120:optim}
 \begin{align}
 & \underset{\pi}{\text{max}}
 & & E^{\pi}[\mathcal{H}(f_{k})-\mathcal{H}(f_{N_{T}})|\mathbf{p}(N_{T}-1),D_{N_{T}-2}] \label{eq120:a} \\
 & \text{s.t.}
 & & E^{\pi}\left[\sum\limits_{t=0}^{N_{T}-1} Z_{t}|\mathbf{p}(N_{T}-1),D_{N_{T}-2} \right] \geq (2\alpha-1)N_{T} \label{eq120:b}
 \end{align}
\end{subequations}
We observe that the constraint \eqref{eq120:b} can also be written as:
\begin{equation}
E^{\pi}\left[\sum\limits_{t=k}^{N_{T}-1} Z_{t}|\mathbf{p}(N_{T}-1),D_{N_{T}-2} \right] \geq (2\alpha-1)N_{T}-\sum\limits_{t=0}^{k-1}Z_{t}
\label{eq121}
\end{equation}
since the pieces of feedback $\{Z_{0},...,Z_{k-1}\}$ already happened. If we manage to reformulate the left hand side of \eqref{eq121} in the fashion of \eqref{eq12:b}, then the problem defined by \eqref{eq120:a} and \eqref{eq121} is solved with the same optimal policy derived for \eqref{eq12:optim}, but with a different $\alpha$. Specifically, we wish the left hand side of \eqref{eq121} to have the form $(2\alpha_{k}-1)(N_{T}-k)$ which by equating the two expressions generates the following $\alpha_{k}$ value:
\begin{equation}
\alpha_{k}=\frac{2\alpha N_{T}-k-\sum\limits_{t=0}^{k-1}Z_{t}}{2(N_{T}-k)}
\label{eq122}
\end{equation}
Therefore, the overall optimal adaptive policy can now be expressed as $\pi^{*}_{0:(N_{T}-1)}=\{\mathbf{p}(0)=C_{0}^{-1}(\alpha_{0}),..,\mathbf{p}(N_{T}-1)=C_{N_{T}-1}^{-1}(\alpha_{N_{T}-1})\}$ where $\alpha_{0}=\alpha$.

\subsection{The Necessity of Exploration}

Here, we need to point out an important issue in AL which was emphasized in our previous work \cite{biban96,biban146}, the necessity of exploration. Reducing the uncertainty of our knowledge about $\mathbf{h}^{*}$ must be performed by approaching this exact value uniformly from all directions. This means that the training samples in an AL process, in this case the power probing vectors, must be diversified and this can be accomplished by choosing hyperplanes in the version space of random direction uniformly. Therefore, we need first to define how to uniformly sample a random direction $\boldsymbol\theta$, where $\boldsymbol\theta$ is a unit vector. This problem is related to the uniform unit hypersphere point picking which has been thoroughly described in \cite{biban96,biban146}. Hence, in order to produce a power vector which represents a hyperplane of random direction, $\mathbf{p}(t)$ must be parallel to a randomly generated $\boldsymbol\theta$, $\mathbf{p}(t)=\beta\boldsymbol\theta$ where $\beta \in \mathbb{R}$, and it must also satisfy $C_{t}(\mathbf{p}(t))=\alpha_{t}$ according to our previous analysis. Essentially, we exploit the degrees of freedom of the design rule in order to introduce exploration into the AL process. In a formal manner, this is expressed using \eqref{eq19} as:
\begin{equation}
\int\limits_{\mathbf{h}\;\beta\boldsymbol\theta^\intercal \leq 1} f_{t}(\mathbf{h})\;dV_{\mathbf{h}}=C_{t}(\beta\boldsymbol\theta)=\alpha_{t}
\label{eq37}.
\end{equation}
At this point, we make use of the Gaussian approximation of each step's posterior pdf which we developed in Section IV with the help of EP. In accordance with that result, $f_{t}(\mathbf{h})$ can be approximated by the normalized version of $\prod\limits_{i=0}^{t-1}\tilde{l}_{i}(\mathbf{h})$ which we denote as $\tilde{f}_{t}(\mathbf{h})$. So, \eqref{eq37} now becomes:
\begin{equation}
\int\limits_{\mathbf{h}\;\beta\boldsymbol\theta^\intercal \leq 1} \tilde{f}_{t}(\mathbf{h})\;dV_{\mathbf{h}}=\alpha_{t}
\label{eq38}
\end{equation}
With the help of the transformation scheme described in Appendix \ref{firstAppendix} and after some processing, we obtain that $\beta=\frac{1}{F^{-1}(\alpha_{t};c_{1},c_{2})}$ where $F^{-1}(.)$ is the inverse cdf of the univariate normal pdf with mean $c_{1}$ and variance $c_{2}$. Furthermore, $c_{1}=\boldsymbol\theta\;\tilde{\bm{\mu}}^\intercal(t)$, where $\tilde{\bm{\mu}}(t)$ is the mean row vector of $\tilde{f}_{t}(\mathbf{h})$, and $c_{2}=\sum\limits_{i=1}^{N}\theta_{i}\boldsymbol\theta(\tilde{\bm{\varSigma}}_{:,i}(t))^\intercal$ , where $\tilde{\bm{\varSigma}}_{:,i}(t)$ is the $i_{th}$ column of the covariance matrix of $\tilde{f}_{t}(\mathbf{h})$. Moreover, all the coordinates of $\mathbf{p}(t)$, which represent power levels, must be non negative, otherwise a new $\boldsymbol\theta$ has to be generated until a valid power vector is produced.


\section{Results}
In this section, we provide simulation results to compare the performance of the CBAL method presented in this work and the constrained versions of the Bayesian AL techniques suggested in \cite{biban146}, the MCMC based Median CPM and the Minimum Volume Ellipsoid (MVE) CPM. The purpose of examining these techniques is to test how fast the analytical and computationally cheap EP based CBAL (EP CBAL) scheme of this paper learns in comparison with the also analytical and computationally cheap MVE CPM, but most importantly compared to the accurate but computationally expensive MCMC based Median CPM. For abbreviation purposes, the two last methods are denoted from here on as MVE CBAL and MCMC CBAL respectively.

The figures of this section show for each AL method the channel estimation error depending on the number of time flops where each time flop is the time period necessary for the CBS to decode the ACK/NACK packet, design the SU probing power vector and probe the PU system. The interference channel gain vector estimation error metric at each time flop is defined as the normalized root-square error $\frac{\lVert \hat{\mathbf{h}}(t)-\mathbf{h}^{*} \rVert}{\lVert \mathbf{h}^{*} \rVert}$ and basically demonstrates the learning efficiency of each method. The estimated interference channel gain vector at each step, $\hat{\mathbf{h}}(t)$, is considered as the $\tilde{\bm{\mu}}(t)$ for the EP CBAL method, the center of the MVE ellipsoid given by every step of the MVE CBAL and the mean calculated at each stage of the MCMC CBAL. The error figure results are obtained as the average of the error metric defined earlier over $100$ SU random topologies, which deliver $100$ random draws of interference channel gain vectors $\mathbf{h}^{*}$.

Moreover, each figure of subsection VI.B is followed by a metric which examines the protection of the PU link quality as the proposed method progresses in time. This can be measured by the time ratio during which the induced interference caused to the PU system is harmless. This is actually the time ratio during which pieces of feedback $Z_{t}=+1$ occur, $\frac{\sum\limits_{t=0}^{N_{T}-1} \mathbf{1}_{\{Z_{t}=+1\}}}{N_{T}}$. This parameter of harmless interference is also averaged over the $100$ SU random topologies to deliver the corresponding average protection metric $\alpha_{sim}$.

\subsection{Simulation Parameters}
As far as the technical parameters of the simulations are concerned, the PU receiver is chosen to normally operate and acknowledge with ACK packets when interference is below $I_{th}=-97\mbox{dBm}$, a limit unknown to the CRN. The examined scenarios consider $N = 5$ and $N=10$ SUs which are dispersed uniformly within a $3\mbox{km}$ range around the PU receiver. The interference channel gains that are unknown to the CRN are assumed to follow an exponential path loss model $g_{i}=\frac{1}{d_{i}^{4}}$, where $d_{i}$ is the distance of the SU\textsubscript{i} from the PU receiver in meters. Additionally, the protection time ratio $\alpha$ takes the following values $\{0.5,\;0.7,\;0.9\}$ where $\alpha=0.5$ basically means that the unconstrained Bayesian AL problem is considered and thus protecting the PU is of no interest. The remaining scenario parameter is the "budget" of $N_{T}$ probing attempts which can also be considered as the pilot time window and it is assumed to be $N_{T}=100$ for the $N = 5$ SU case and $N_{T}=200$ for the $N=10$ SU case. Additionally, a practical consideration which must be taken into account is the necessary number of samples for the MCMC CBAL method to be accurate, which for $N=5$ dimensions is $N_{r}=20000$ and for $N=10$ dimensions is $N_{r}=150000$. This sample number increases exponentially depending on the learning problem dimensions and it is the great disadvantage of this numerical tool. Our proposed EP based solution tackles exactly this issue and produces posterior pdf approximations of high accuracy with low computational burden exactly because it is analytical and not numerical. 

\subsection{Estimation Performance of the Constrained Bayesian AL Method}

\begin{figure}[!h]
\centering
\includegraphics[scale=0.68, trim=17 0 0 10, clip]{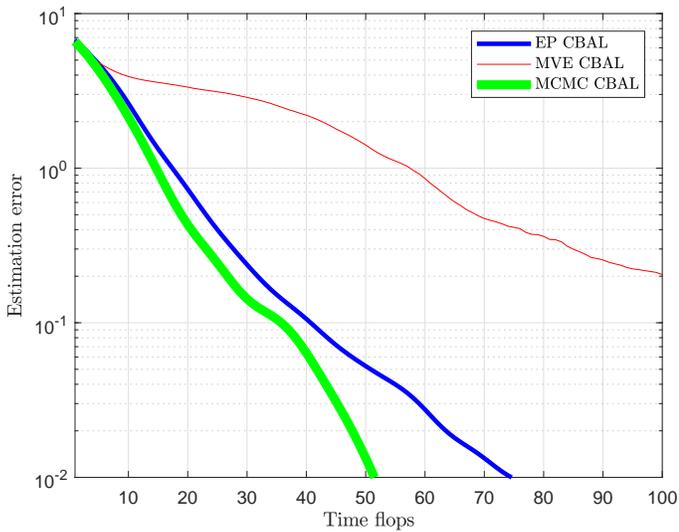}
\caption{Interference channel gain vector estimation error progress vs time of the CBAL methods for $\alpha=0.5$ and $N=5$ SUs}
\label{fig2}
\end{figure}

\begin{figure}[!h]
\centering
\includegraphics[scale=0.68, trim=17 0 0 10, clip]{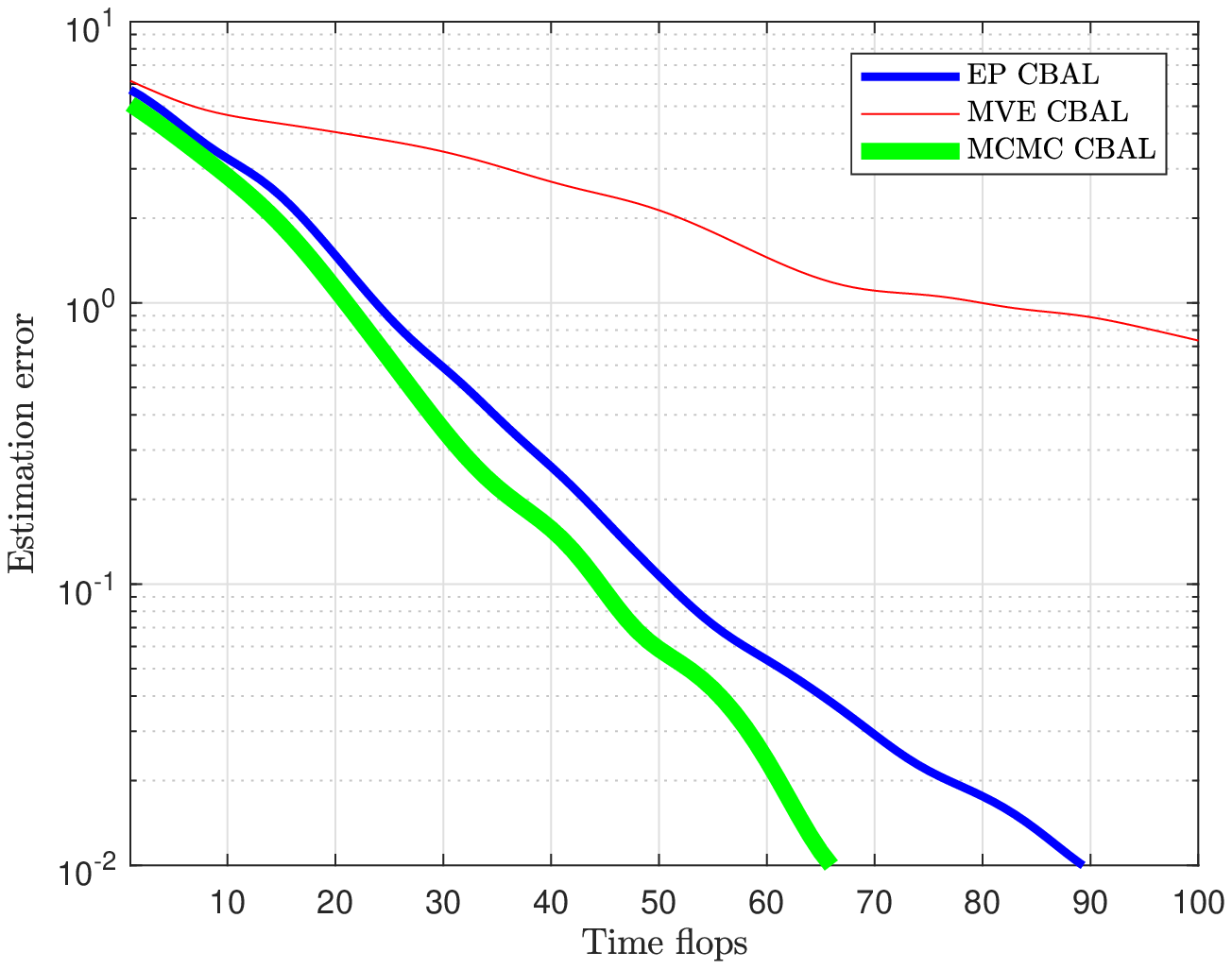}
\caption{Interference channel gain vector estimation error progress vs time of the CBAL methods for $\alpha=0.7$ and $N=5$ SUs}
\label{fig3}
\end{figure}

\begin{figure}[!h]
\centering
\includegraphics[scale=0.68, trim=17 0 0 10, clip]{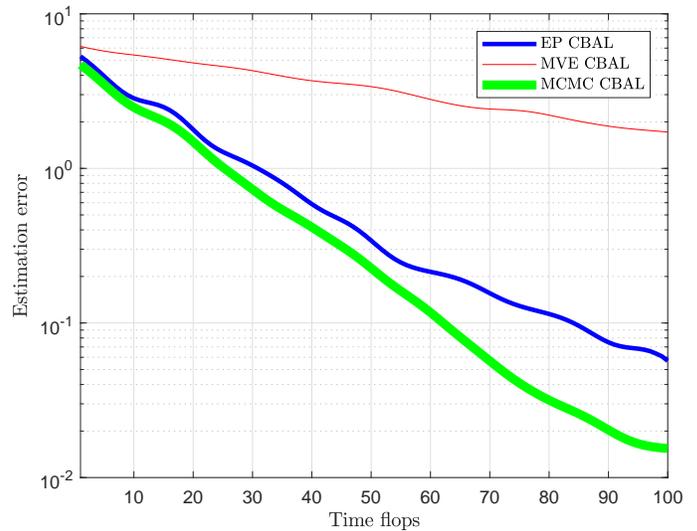}
\caption{Interference channel gain vector estimation error progress vs time of the CBAL methods for $\alpha=0.9$ and $N=5$ SUs}
\label{fig4}
\end{figure}

Initially, let us see in Fig. \ref{fig2}, \ref{fig3} and \ref{fig4} the performance of all the considered CBAL techniques for $N=5$ SUs. At first, it can be clearly seen that as $\alpha$ is increased, more probing attempts are required to correctly estimate $\mathbf{h}^{*}$. Furthermore, the MCMC CBAL scheme outperforms in speed both EP CBAL and MVE CBAL. More specifically, in the case of $\alpha=0.5$, Fig. \ref{fig2}, for an estimation error $1\%$, the MCMC CBAL and the EP CBAL schemes converge in $51$ and $72$ time flops respectively, while the MVC CBAL hardly reaches a  $20\%$ error at $100$ time flops. For $\alpha=0.7$, as it can be seen in Fig. \ref{fig3}, the corresponding required time flops for an estimation error $1\%$ are $65$ and $88$ for the MCMC CBAL and the EP CBAL techniques, whereas the MVE CBAL scheme exhibits a severely deteriorated convergence. For $\alpha=0.9$, convergence worsens even further for all schemes as shown in Fig. \ref{fig4}, where after $100$ probing attempts, the estimation errors are $1.7\%$ and $6\%$ for the MCMC CBAL and the EP CBAL techniques. These results prove that as the design parameter of PU protection $\alpha$ increases, the CBS designs less harmful for the PU system probing power vectors, but also less informative about $\mathbf{h}^{*}$.

As far as the $\alpha_{sim}$ metric for these three cases is concerned, for $\alpha=0.5$, $\alpha=0.7$ and $\alpha=0.9$, the resulting $\alpha_{sim}$ values for the MCMC CBAL technique are $\alpha_{sim}=0.5$, $\alpha_{sim}=0.72$ and $\alpha_{sim}=0.91$ respectively and for the EP CBAL scheme are $\alpha_{sim}=0.49$, $\alpha_{sim}=0.68$ and $\alpha_{sim}=0.87$ respectively. The small differences between the target values of the protection time ratio, $\alpha$, and the simulated ones, $\alpha_{sim}$, appear because of the inaccurate estimation of the each step posterior pdf using either MCMC's or the EP. Even though EP is a very accurate, sophisticated and fast method for density estimation, the approximated posterior pdf's still have some deviation from the real ones. This results in computing power vectors which satisfy \eqref{eq38} but not its exact version, \eqref{eq37}. Similar but slightly smaller deviations are observed for the $\alpha_{sim}$ values of the MCMC CBAL technique. As far as the corresponding $\alpha_{sim}$ of the MVE CBAL method are concerned, these are $\alpha_{sim}=0.56$, $\alpha_{sim}=0.63$ and $\alpha_{sim}=0.8$ and cannot be considered adequately close to the design $\alpha$ values.

\begin{figure}[!h]
\centering
\includegraphics[scale=0.68, trim=17 0 0 10, clip]{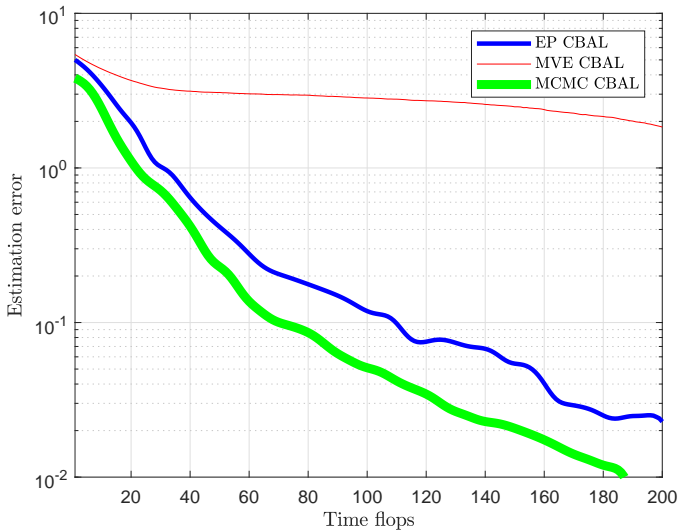}
\caption{Interference channel gain vector estimation error progress vs time of the CBAL methods for $\alpha=0.7$ and $N=10$ SUs}
\label{fig5}
\end{figure}

Next, we examine for $N=10$ SUs and designed protection time ratio $\alpha=0.7$ the performance of all the techniques which is illustrated in Fig. \ref{fig5}. After $N_{T}=186$ time flops, the $\mathbf{h}^{*}$ estimation error for the MCMC CBAL method is $1\%$, while after $N_{T}=200$ time flops the estimation error for the EP CBAL scheme is $2.5\%$ and for the MVE CBAL technique it is again beyond comparison. The respective simulated protection time ratios are $\alpha_{sim}=0.7$, $\alpha_{sim}=0.67$ and $\alpha_{sim}=0.78$. The reason for checking the learning efficiencies for $N=10$ SUs is first to observe their behavior when the learning problem dimensions grow. We observe by comparing the results of Fig. \ref{fig3} and \ref{fig5} that the convergence time for all methods increases which is reasonable, because a greater number of parameters, the interference channel gains, is being sought which demands more probing attempts. Second, we wish to show that the learning performance of the EP CBAL scheme does not diverge from the one of the MCMC CBAL as the problem dimensions grow. Subsequently, this proves that the EP posterior pdf approximation does not deteriorate as $N$, or the number of SUs, increases. More specifically, we observe by comparing the results of Fig. \ref{fig3} and \ref{fig5} that the convergence time of the EP CBAL method for an estimation error of $2.5\%$ increases from $72$ time flops in Fig. \ref{fig3} to $200$ in Fig. \ref{fig5}. Hence, we could empirically claim that the convergence rate of the proposed method in this paper depending on the number of SUs, $N$, is of slightly higher order than $\mathcal{O}(N\log_{2}N)$.

As mentioned earlier, the main purpose for comparing the EP CBAL method with the MCMC CBAL and the MVE CBAL ones is to compare their learning convergence rates. Obviously, the MCMC CBAL technique outperforms the proposed method of this paper. Nevertheless, this comes with a heavy penalty. The MCMC tool requires the generation of exhaustively many random samples in the $\mathbf{h}$ space at each time step. The number of these samples grows exponentially with the problem dimensions, $N$, and this makes the MCMC CBAL scheme an unrealistic choice for a CBS where all these computations take place in order to design the SU probing power vectors. This problem worsens if the CBS has limited computational capabilities. Our proposed analytical scheme, the EP CBAL, tackles exactly this issue. It offers a computationally cheap and accurate alternative to the MCMC based AL scheme which exhibits slightly slower convergence. Thus, the EP CBAL manages to combine the benefits of the previously developed methods in \cite{biban146}, a high accuracy of the posterior pdf computation comparable to that of the MCMC tool which subsequently delivers fast learning convergence rates and the analytical and therefore fast computation of the posterior pdf at each time step which similarly to the MVE CBAL does not burden the CBS.


\section{Conclusions and Future Work}
In this paper, we proposed a sequential probing method in order for a centralized CRN to learn fast the PU interference constraint using the ACK/NACK PU feedback while constraining the number of PU outage events. This problem was formulated within the Constrained DP framework and its optimal solution policy was implemented with the help of an advanced, fast and accurate Bayesian Learning technique, the EP, which was for the first time developed analytically without independence assumptions about the latent variables. The performance of this method was demonstrated through numerical simulations in static channel scenarios for interference channel gain learning and compared to constrained versions of Bayesian AL schemes we earlier developed in \cite{biban146}. Additionally, we confirmed that the simulated PU protection metric $\alpha_{sim}$, which is basically the complementary of the induced PU outage time ratio, is satisfactorily close to the target, or design, PU protection time ratio $\alpha$.  

As part of our future work, the same idea of CBAL for interference channel gain learning will be studied in the case of uncertain ACK/NACK feedback which occurs under low SINR conditions of the sensed PU signal on the CRN side. Within the AL framework, we also plan to develop variations of our current methods which will be suitable for a decentralized CRN structure with a message passing mechanism between the SUs. This subject has been studied in collaborative cognitive radar scenarios but without using sophisticated learning mechanisms. Such AL methods are closely related to decentralized learning schemes and could tackle issues like scalability. Additionally, asynchronous decentralized learning schemes could be investigated which are of great practical importance especially in communication systems.


\appendices

\section{Moments of a one side truncated multivariate normal pdf}
\label{firstAppendix}

Assuming a multivariate normal pdf $\mathcal{N}(\mathbf{x};\bm{\mu}_{\mathbf{x}},\bm{\varSigma}_{\mathbf{x}})$ of $N$ dimensions and a halfspace indicator function:
\begin{align}
g(\mathbf{x})=
\left\{
  \begin{array}{cc}
   1 & \mbox{if $\mathbf{a}\;\mathbf{x}^\intercal\leq b$}\\
   0 & \mbox{if $\mathbf{a}\;\mathbf{x}^\intercal> b$}
  \end{array}
  \right.
\label{appA:eq1}
\end{align}
where $\mathbf{a}\;\mathbf{x}^\intercal= b$ is the hyperplane limit of this halfspace and $\mathbf{a}$ and $\mathbf{x}$ are row vectors, then $h(\mathbf{x})=g(\mathbf{x})\;\mathcal{N}(\mathbf{x};\bm{\mu}_{\mathbf{x}},\bm{\varSigma}_{\mathbf{x}})$ is an un-normalized one side truncated multivariate normal pdf. Next, we determine the $0_{th}$, $1_{st}$ and $2_{nd}$ moments of $h(\mathbf{x})$, $q$, $\mathbf{q}$ and $\mathbf{Q}$, based on the moment related integrals, $c$, $\mathbf{c}$ and $\mathbf{C}$:

\begin{equation}
c=\int\limits_{\mathbb{R}^{N}} h(\mathbf{x})\;dV_{\mathbf{x}}=\int\limits_{\mathbf{a}\;\mathbf{x}^\intercal\leq b} \mathcal{N}(\mathbf{x};\bm{\mu}_{\mathbf{x}},\bm{\varSigma}_{\mathbf{x}})\;dV_{\mathbf{x}}
\label{appA:eq5a}
\end{equation}

\begin{equation}
\mathbf{c}=\int\limits_{\mathbb{R}^{N}} \mathbf{x}\;h(\mathbf{x})\;dV_{\mathbf{x}}=\int\limits_{\mathbf{a}\;\mathbf{x}^\intercal\leq b} \mathbf{x}\;\mathcal{N}(\mathbf{x};\bm{\mu}_{\mathbf{x}},\bm{\varSigma}_{\mathbf{x}})\;dV_{\mathbf{x}}
\label{appA:eq5b}
\end{equation}

\begin{equation}
\mathbf{C}=\int\limits_{\mathbb{R}^{N}} \mathbf{x}^\intercal\mathbf{x}\;h(\mathbf{x})\;dV_{\mathbf{x}}=\int\limits_{\mathbf{a}\;\mathbf{x}^\intercal\leq b} \mathbf{x}^\intercal\mathbf{x}\;\mathcal{N}(\mathbf{x};\bm{\mu}_{\mathbf{x}},\bm{\varSigma}_{\mathbf{x}})\;dV_{\mathbf{x}}
\label{appA:eq5c}.
\end{equation}
Note that $c$ is a constant which represents the mass or the normalization factor of $h(\mathbf{x})$, $\mathbf{c}$ is a vector of integrals and $\mathbf{C}$ is a matrix of integrals. The moments can be written as $q=c$, $\mathbf{q}=\frac{\mathbf{c}}{c}$ and $\mathbf{Q}=\frac{\mathbf{C}}{c}-\mathbf{q}^\intercal \mathbf{q}$. The problem of computing these moments lies on the computation of the integrals in \eqref{appA:eq5a}, \eqref{appA:eq5b} and \eqref{appA:eq5c}.

Now, if we define an $N \times N$ transformation matrix $T$ such as:
\begin{equation}
T=\begin{bmatrix}
    a_{1} & 0 & 0 & \dots & 0 & 0 \\
    a_{2} & 1 & 0 & \dots & 0 & 0 \\
    a_{3} & 0 & 1 & \dots & 0 & 0 \\
    \vdots & \vdots & \vdots & \ddots & \vdots & \vdots \\
    a_{N-1} & 0 & 0 & \dots & 1 & 0 \\
    a_{N} & 0 & 0 & \dots & 0 & 1
\end{bmatrix}
\label{appA:eq2}
\end{equation}
and determine a new random variable $\mathbf{y}=\mathbf{x}\;T$, then $\mathbf{y}$ will also be normally distributed, $\mathbf{y} \sim \mathcal{N}(\mathbf{y};\bm{\mu}_{\mathbf{y}},\bm{\varSigma}_{\mathbf{y}})$, where:
\begin{equation}
\bm{\mu}_{\mathbf{y}}=\bm{\mu}_{\mathbf{x}}\;T
\label{appA:eq3}
\end{equation}
and
\begin{equation}
\bm{\varSigma}_{\mathbf{y}}=T^\intercal\bm{\varSigma}_{\mathbf{x}}T
\label{appA:eq4}.
\end{equation}
This helps us transform the integrals in \eqref{appA:eq5a}, \eqref{appA:eq5b} and \eqref{appA:eq5c} by using the change-of-variables technique. The Jacobian matrix $J_{\mathbf{x}\rightarrow\mathbf{y}}$ is actually equal to $T^\intercal$, hence the infinitesimal volume $dV_{\mathbf{x}}$ can be rewritten as $\frac{dV_{\mathbf{y}}}{|det(T^\intercal)|}$ or $\frac{dV_{\mathbf{y}}}{|det(T)|}$. Using this and changing the integral limits delivers the following for \eqref{appA:eq5a}:
\begin{align}
& c=\int\limits_{\mathbf{a}\;\mathbf{x}^\intercal\leq b} \mathcal{N}(\mathbf{x};\bm{\mu}_{\mathbf{x}},\bm{\varSigma}_{\mathbf{x}})\;dV_{\mathbf{x}}=\nonumber \\
& \int\limits_{y_{1}\leq b} \int_{-\infty}^{\infty}...\int_{-\infty}^{\infty} \mathcal{N}(\mathbf{x};\bm{\mu}_{\mathbf{x}},\bm{\varSigma}_{\mathbf{x}})\;\frac{dV_{\mathbf{y}}}{|det(T)|}=\nonumber \\
& \int_{-\infty}^{y_{1}= b} \int_{-\infty}^{\infty}...\int_{-\infty}^{\infty} \frac{\mathcal{N}(\mathbf{x};\bm{\mu}_{\mathbf{x}},\bm{\varSigma}_{\mathbf{x}})}{|det(T)|}\;dV_{\mathbf{y}}=\nonumber \\
& \int_{-\infty}^{y_{1}= b} \int_{-\infty}^{\infty}...\int_{-\infty}^{\infty} \mathcal{N}(\mathbf{y};\bm{\mu}_{\mathbf{y}},\bm{\varSigma}_{\mathbf{y}})\;dV_{\mathbf{y}}
\label{appA:eq7a}
\end{align}
where in the last line we used the relation of the two pdf's of the random variables $\mathbf{x}$ and $\mathbf{y}$. Similarly, for $\mathbf{c}$ and $\mathbf{C}$, we have:
\begin{equation}
\mathbf{c}=\left(\int_{-\infty}^{y_{1}= b} \int_{-\infty}^{\infty}...\int_{-\infty}^{\infty} \mathbf{y}\;\mathcal{N}(\mathbf{y};\bm{\mu}_{\mathbf{y}},\bm{\varSigma}_{\mathbf{y}})\;dV_{\mathbf{y}}\right)T^{-1}
\label{appA:eq7b}
\end{equation}
and
\begin{align}
& \mathbf{C}=\nonumber \\
& (T^{-1})^\intercal\left(\int_{-\infty}^{y_{1}= b} \int_{-\infty}^{\infty}...\int_{-\infty}^{\infty} \mathbf{y}^\intercal\mathbf{y}\;\mathcal{N}(\mathbf{y};\bm{\mu}_{\mathbf{y}},\bm{\varSigma}_{\mathbf{y}})\;dV_{\mathbf{y}}\right)T^{-1}
\label{appA:eq7c}.
\end{align}

Consequently, the problem of calculating the moments of a one side truncated multivariate Gaussian pdf has been transformed into calculating the moments of another one side truncated multivariate Gaussian pdf where the truncation occurs vertically to the axis $y_{1}y_{1}'$. This is the study object of Appendix \ref{secondAppendix}.

\section{Moments of a one vertical side truncated multivariate normal pdf}
\label{secondAppendix}

In this section, we elaborate on the moments of one vertical side truncated multivariate normal pdf's. In the statistics literature, the truncation subject has been extensively investigated using many kinds of truncations, such as box-like and elliptical ones. Here, we present a simplified case of calculating the moments of a doubly truncated multivariate normal pdf recently studied in \cite{biban191} and which actually concerns a hyper-rectangle truncation. The simplification introduced here will lead us to computing the moments of the one vertical side truncated multivariate normal pdf. Assuming a multivariate normal pdf $\mathcal{N}(\mathbf{x};\bm{\mu},\bm{\varSigma})$ in $N$ dimensions and a hyper-rectangle defined by the inequalities $a_{i}\leq x_{i} \leq b_{i}$ for $i=1,..,N$, the authors of \cite{biban191} managed to find simple recursive relations for the moment related integrals and therefore allow the fast computation of doubly truncated multivariate normal pdf's moments.

More specifically, if $\mathbf{a}=[a_{1},...,a_{N}]$ and $\mathbf{b}=[b_{1},...,b_{N}]$, then $L_{\mathbf{k}}(\mathbf{a},\mathbf{b};\bm{\mu},\bm{\varSigma})$ is the integral defined as:
\begin{equation}
L_{\mathbf{k}}(\mathbf{a},\mathbf{b};\bm{\mu},\bm{\varSigma})=\int_{a_{1}}^{b_{1}} ...\int_{a_{N}}^{b_{N}} \mathbf{x}^{\mathbf{k}}\;\mathcal{N}(\mathbf{x};\bm{\mu},\bm{\varSigma})\;dV_{\mathbf{x}}
\label{appB:eq1}
\end{equation}
where $\mathbf{x}^{\mathbf{k}}$ stands for $x_{1}^{k_{1}}\cdot...\cdot x_{N}^{k_{N}}$. For example, if we wish to compute the integral $\int_{a_{1}}^{b_{1}} ...\int_{a_{4}}^{b_{4}} x_{1}x_{3}\;\mathcal{N}(\mathbf{x};\bm{\mu},\bm{\varSigma})\;dV_{\mathbf{x}}$ for $N=4$, then $\mathbf{k}=[1,0,1,0]$. Additionally, we denote by $\mathbf{r}_{(i)}$ a row vector $\mathbf{r}$ with its $i_{th}$ element removed, by $\mathbf{R}_{i,(j)}$ the $i_{th}$ row of a matrix $\mathbf{R}$ with its $j_{th}$ element removed, by $\mathbf{R}_{(i),j}$ the $j_{th}$ column of a matrix $\mathbf{R}$ with its $i_{th}$ element removed and by $\mathbf{R}_{(i),(j)}$ a matrix $\mathbf{R}$ with its $i_{th}$ row and $j_{th}$ column removed. In \cite{biban191}, it is shown that if we let $\mathbf{e}_{i}$ denote an $N$-dimensional row vector with its $i_{th}$ element equal to one and zeros otherwise, then:
\begin{equation}
L_{\mathbf{k}+\mathbf{e}_{i}}(\mathbf{a},\mathbf{b};\bm{\mu},\bm{\varSigma})=\mu_{i} L_{\mathbf{k}}(\mathbf{a},\mathbf{b};\bm{\mu},\bm{\varSigma})+\mathbf{e}_{i}\bm{\varSigma}\mathbf{c}_{\mathbf{k}}^\intercal
\label{appB:eq2}
\end{equation}
where $\mathbf{c}_{\mathbf{k}}$ is an $N$-dimensional row vector with its $j_{th}$ element equal to:
\begin{align}
& \mathbf{c}_{\mathbf{k},j}=k_{j}\;L_{\mathbf{k}-\mathbf{e}_{j}}(\mathbf{a},\mathbf{b};\bm{\mu},\bm{\varSigma})+\nonumber \\
& +a_{j}^{k_{j}}\;\mathcal{N}(a_{j};\mu_{j},\bm{\varSigma}_{j,j})\;L_{\mathbf{k}_{(j)}}(\mathbf{a}_{(j)},\mathbf{b}_{(j)};\tilde{\bm{\mu}}_{j}^{\mathbf{a}},\tilde{\bm{\varSigma}}_{j})+\nonumber \\
&+b_{j}^{k_{j}}\;\mathcal{N}(b_{j};\mu_{j},\bm{\varSigma}_{j,j})\;L_{\mathbf{k}_{(j)}}(\mathbf{a}_{(j)},\mathbf{b}_{(j)};\tilde{\bm{\mu}}_{j}^{\mathbf{b}},\tilde{\bm{\varSigma}}_{j})
\label{appB:eq3}
\end{align}
and
\begin{equation}
\tilde{\bm{\mu}}_{j}^{\mathbf{a}}=\bm{\mu}_{(j)}+\bm{\varSigma}_{j,(j)}\frac{a_{j}-\mu_{j}}{\bm{\varSigma}_{j,j}}
\label{appB:eq4}
\end{equation}
\begin{equation}
\tilde{\bm{\mu}}_{j}^{\mathbf{b}}=\bm{\mu}_{(j)}+\bm{\varSigma}_{j,(j)}\frac{b_{j}-\mu_{j}}{\bm{\varSigma}_{j,j}}
\label{appB:eq5}
\end{equation}
\begin{equation}
\tilde{\bm{\varSigma}}_{j}=\bm{\varSigma}_{(j),(j)}-\frac{\bm{\varSigma}_{(j),j}\;\bm{\varSigma}_{j,(j)}}{\bm{\varSigma}_{j,j}}
\label{appB:eq6}.
\end{equation}
Hence, if we intend to obtain the integrals $\int_{a_{1}}^{b_{1}} ...\int_{a_{N}}^{b_{N}} x_{m}\;\mathcal{N}(\mathbf{x};\bm{\mu},\bm{\varSigma})\;dV_{\mathbf{x}}$ for $m=1,...,N$ and calculate the mean of a doubly truncated multivariate normal pdf, then we should set $\mathbf{k}=\mathbf{0}$ and $\mathbf{e}_{i}=\mathbf{e}_{m}$ in \eqref{appB:eq2}. Next, we should divide the results with the normalization constant of the truncated Gaussian $\int_{a_{1}}^{b_{1}} ...\int_{a_{N}}^{b_{N}} \mathcal{N}(\mathbf{x};\bm{\mu},\bm{\varSigma})\;dV_{\mathbf{x}}$, which in \cite{biban191} is calculated using the inclusion-exclusion principle, a combinatorics technique. Similarly, for the $2_{nd}$ order moment, we are interested in computing integrals of the form $\int_{a_{1}}^{b_{1}} ...\int_{a_{N}}^{b_{N}} x_{m}x_{n}\;\mathcal{N}(\mathbf{x};\bm{\mu},\bm{\varSigma})\;dV_{\mathbf{x}}$ for $m=1,...,N$ and $n=1,...,N$ which can be acquired by setting $\mathbf{k}=\mathbf{e}_{m}$ and $\mathbf{e}_{i}=\mathbf{e}_{n}$ in \eqref{appB:eq2}.

Now, if we let $\mathbf{a}$ and $\mathbf{b}$, which define the box-like truncation, be respectively $[-\infty,...,-\infty]$ and $[b_{1},\infty,...,\infty]$, then the aforementioned recursive relations concern the moments of a one vertical side truncated multivariate normal pdf, where the cutting hyperplane is $x_{1}=b_{1}$ and the hyper-rectangle is now the halfspace $x_{1}\leq b_{1}$. The relations \eqref{appB:eq2}, \eqref{appB:eq3}, \eqref{appB:eq4}, \eqref{appB:eq5} and \eqref{appB:eq6} are simplified and moreover we have the benefit of not using the inclusion-exclusion principle, which for large $N$ can be computationally demanding, for the calculation of the mass of the truncated $\mathcal{N}(\mathbf{x};\bm{\mu},\bm{\varSigma})$. This happens because $\int_{-\infty}^{b_{1}} \int_{-\infty}^{\infty}...\int_{-\infty}^{\infty} \mathcal{N}(\mathbf{x};\bm{\mu},\bm{\varSigma})\;dV_{\mathbf{x}}$ is actually equal to $\int_{-\infty}^{b_{1}} \mathcal{N}(x_{1};\mu_{1},\bm{\varSigma}_{1,1})\;dx_{1}$.


\bibliographystyle{IEEEtran}
\bibliography{Bibliografia}




\end{document}